\documentclass[preprint2]{aastex}
\input psfig.sty

\newcommand{\beq}{\begin{equation}}
\newcommand{\eeq}{\end{equation}}
\newcommand{\rsht}{{MASTER}}
\newcommand{\tC}{\widetilde{C}}
\newcommand{\tN}{\widetilde{N}}
\renewcommand{\r}{{\bf{r}}}
\newcommand{\n}{{\bf{n}}}
\renewcommand{\k}{{\bf{k}}}
\newcommand{\fsky}{{f_{\rm sky}}}
\newcommand{\fzero}{{F^{(0)}}}
\newcommand{\fzerol}{{F^{(0)}_{\ell}}}
\newcommand{\npix}{{N_{\rm pix}}}
\newcommand{\nbins}{{n_{\rm bins}}}
\newcommand{\lmax}{{\ell_{\rm max}}}
\newcommand{\nmc}{{N_{\rm MC}}}
\newcommand{\nmcs}{{N_{\rm MC}^{\rm (s)}}}
\newcommand{\nmcn}{{N_{\rm MC}^{\rm (n)}}}
\newcommand{\nmcsn}{{N_{\rm MC}^{\rm (s+n)}}}
\newcommand{\ntau}{{N_{\tau}}}
\newcommand{\nfft}{{N_{\rm FFT}}}
\newcommand{\Cltheory}{{C_\ell^{\rm th}}}
\newcommand{\Ctheory}{{C^{\rm th}}}

\newcommand{\VEV}[1]{\langle#1\rangle}

\newcommand{\bldb}{{Boom-LDB}}
\newcommand{\wjjj}[6]
{{
\left( 
\begin{array}{lcr} #1 & #2 & #3 \\#4 & #5 & #6 \end{array}
\right) 
}}

\def\la{\mathrel{\mathchoice {\vcenter{\offinterlineskip\halign{\hfil
$\displaystyle##$\hfil\cr<\cr\noalign{\vskip1.5pt}\sim\cr}}}
{\vcenter{\offinterlineskip\halign{\hfil$\textstyle##$\hfil\cr<\cr
\noalign{\vskip1.0pt}\sim\cr}}}
{\vcenter{\offinterlineskip\halign{\hfil$\scriptstyle##$\hfil\cr<\cr
\noalign{\vskip0.5pt}\sim\cr}}}
{\vcenter{\offinterlineskip\halign{\hfil$\scriptscriptstyle##$\hfil
\cr<\cr\noalign{\vskip0.5pt}\sim\cr}}}}}

\shorttitle{MASTER}
\shortauthors{Hivon et al}
\begin{document}
\title{MASTER of the CMB Anisotropy Power Spectrum:
A Fast Method for Statistical Analysis of Large and Complex CMB Data Sets}

\author{Eric Hivon\altaffilmark{{1,2}}, 
Krzysztof M. G\'orski\altaffilmark{3,4}, 
C. Barth Netterfield\altaffilmark{5},\\
Brendan P. Crill\altaffilmark{1}, 
Simon Prunet\altaffilmark{6},
Frode Hansen\altaffilmark{7}}

\affil{
$^{1}$ Observational Cosmology, MS 59-33, Caltech, Pasadena, CA 91125 \\
$^{2}$ IPAC, MS 100-22, Caltech, Pasadena, CA 91125 \\
$^{3}$ European Southern Observatory, Garching bei M{\"u}nchen, Germany\\
$^{4}$ Warsaw University Observatory, Warsaw, Poland \\
$^{5}$ Dept. of Phys. and Astron., U. of Toronto, 60 St George St,
Toronto, Ontario, M5S 3H8, Canada \\
$^{6}$ CITA, University of Toronto, 60 St George St, Toronto, Ontario, M5S 3H8, Canada\\
$^{7}$ MPA, Garching, Germany}

\begin{abstract} 
We describe a fast and accurate method for estimation of
the cosmic microwave background (CMB) anisotropy angular
power spectrum --- $\bf M$onte Carlo $\bf A$podised $\bf S$pherical
$\bf T$ransform $\bf E$stimato$\bf R$.
Originally devised for use in the interpretation of the
Boomerang experimental data, 
MASTER  is both 
a computationally efficient method suitable for use with the 
currently available CMB data sets (already large in size, despite
covering small fractions of the sky, and affected by inhomogeneous and
correlated noise),
and a very promising  application for the analysis of  very large
future CMB satellite mission products.
\vspace{1cm}
%It is  based on a direct Spherical
%Harmonics Transforms (SHT) of the observed map. 
%%The mode-mode coupling of the power spectrum
%induced by the incomplete sky coverage 
%can be described analytically as a function
%of the sky window SHT and corrected for in order to get an unbiased estimate o%f
%$C_\ell$. 
%The contribution of the instrumental noise to the derived power
%spectrum as well as the effect of any alteration of the data stream or  the
%map introduced  in the process of data analysis  
%are measured  in Monte Carlo (MC) simulations
%and can be removed or corrected for in the final power spectrum.
%The computation time is generally dominated by the map making process which
%scales like $ {\mathcal{O}}(\nmc \ntau \log \ntau)$, where $\ntau$ is the
%number of time samples, and  $\nmc$ is the total number 
%of Monte Carlo simulations
%%required, which is of the order of a few hundreds.
%We show that this method renders an unbiased estimate of $C_\ell$ when
%tested on simulations featuring the same scanning strategy and noise properties
%as those of the long duration flight of Boomerang (Netterfield et al, 2001, and
%references therein).
\end{abstract}

\section{Introduction}
During the past decade
since the ground-breaking discovery of the cosmic microwave background 
radiation anisotropy by the {\it COBE} satellite (Smoot et al. 1992),
numerous  successful measurements of  microwave sky structures 
have provided us with the data for powerful tests of the  current 
cosmological paradigm, 
and created an unprecedented opportunity 
to estimate key  parameters of the candidate theoretical 
models of the Universe.

Recent ground-based and balloon-borne experiments with improved  sky
coverage, angular resolution, and
noise performance (see de Bernardis et al, 2000, Hanany et al, 2000,
Padin et al, 2000,
Jaffe et al, 2001, Lee et al, 2001, Halverson et al, 2001, Pryke et al, 2001 and references therein 
for some of the most recent experiments and
their interpretation) have both
given us a taste of what future satellite missions 
MAP\footnote{Microwave Anisotropy Probe, http://map.gsfc.nasa.gov/} 
and Planck\footnote{http://astro.estec.esa.nl/Planck/} should
accomplish, and
revealed the growing challenges that we will have to meet in the
analysis of the forthcoming CMB data sets.

In the currently favoured structure formation model of inflation 
induced, Gaussian distributed, curvature 
perturbations
all the statistical information contained in a CMB map
can be summarised in its angular power spectrum $C_\ell$. 
General maximum likelihood methods for extracting $C_\ell$ 
from a $\npix$-pixel map
with non uniform coverage and correlated noise (G\'orski 1994, Bond 1995,
Tegmark \& Bunn 1995,
G\'orski 1997,
Bond, Jaffe, \& Knox 1998,
Borrill 1999b) involve computations
of complexity $\sim \npix^3$,
and become prohibitively CPU expensive for the $\npix >
10^4$  maps produced by current experiments.
With the presently anticipated computer performance
such methods appear totally impractical
for application to the $\npix > 10^6$ maps 
expected from the future space missions (Borrill 1999a). 
Hence, there is a well recognised need for 
faster, more economical, and accurate $C_\ell$ extraction  methods,
which should enable a correct comological interpretion of the 
CMB anisotropy observations.

In this paper we introduce and discuss a new method
for fast estimation  of the CMB anisotropy angular 
power spectrum from fluctuations observed on a limited area of the sky.
This method is  based on a direct spherical
harmonic transform (SHT) of the available  map and 
allows one to incorporate 
a description of 
the particular properties of a given CMB experiment, including the
survey geometry, scanning strategy, instrumental noise
behaviour, and possible non-gaussian and/or 
non-stationary events which can occur during the data acquisition.
The estimated power spectrum is affected by the
unwanted contribution of the instrumental noise 
and the effects of any necessary alteration of either the recorded data stream
(such as high pass filtering) or the raw map of the observed region of
the sky, which are introduced during the data analysis.
These effects
are calibrated in Monte Carlo (MC) simulations
of the modeled observation and analysis stage of the experiment 
and can then be  removed, or corrected for in
the estimated  power spectrum. 
The harmonic mode-mode coupling induced by the incomplete
sky coverage is described analytically by the SHT of the sky window 
and corrected for in order to obtain an unbiased estimate of
the $C_\ell$. 
Hereafter we refer to this method with an acronym MASTER (
Monte Carlo Apodised Spherical Transform EstimatoR).

Netterfield et al. (2001) described an application of this method in
the extraction of the CMB angular power spectrum $C_\ell$, for 
$75 < \ell < 1025$, from the sky map (analysed region comprised 
$\sim $1.8\% of the sky
covered with 57000 pixels of $7'$ size) 
made by  coadding four frequency  channel data 
of the 1998/99 Antarctic 
long duration flight of the Boomerang experiment (\bldb).
The first derivation of the CMB anisotropy spectrum from the same data
(de Bernardis et al. 2000) involved the MADCAP method (Borrill 1999b)
applied to a smaller subset of the data 
(one frequency channel, $\sim $1\% of the sky
covered with $\sim 8000$ pixels of $14'$ size). The MADCAP approach 
is too CPU intensive for repeated applications to the new, enlarged 
subset of the Boomerang data, and, hence, the  MASTER  approach was
the method of choice for extraction of the high-$\ell$
angular power spectrum of the CMB anisotropy.
% and the 
% pertinent cosmological results.

Other fast methods have recently been proposed for estimation of the
angular spectrum of the CMB anistropy.
Szapudi et al. (2001) advocate the use of the 2-point correlation 
function for extraction of  the angular
power spectrum from the CMB maps. The computational demands of this
method
scale quadratically, $\sim \npix^2$, with the size of data set
(that may be improved to $\sim \npix \log\npix$). 
In the same way as in the case of 
\rsht, the effects of the
noise and  correlations of the derived $C_\ell$-s 
are quantified by Monte Carlo
simulations (although the demonstrated applications involved only the
case of a
uniform white noise).
Dor\'e, Knox \& Peel (2001b) proposed a hierarchical 
implementation of the usual
quadratic $C_\ell$ estimator with a computational  scaling  
proportional to $\npix^2$, that may be
reduced to $\sim \npix$ (with a large prefactor) 
at the price of additional approximations.

Experiment specific techniques have also been proposed :
Oh, Spergel, and Hinshaw (1999) 
described a fast power spectrum 
extraction
technique designed for usage with the MAP satellite data.
Their method scales like $\sim \npix^2$ with the size of the
pixellised map, and takes advantage of uncorrelated pixel noise with 
approximate axisymmetric distribution on the sky.  
Wandelt (2000) advocates the use of the set of  rings
as a
compressed form of the Planck data set from which to extract optimally the
$C_\ell$-s  in the presence of correlated noise. The applicability of
this approach
is limited by its assumption of the 
the symmetry of the scanning strategy.

This paper is organised as follows:
In section \ref{section:TODtoCl} we describe how a data stream of observations
is reduced to a CMB fluctuation map, 
and how the  angular pseudo power spectrum $\tC_\ell$ 
is extracted from such a  map by SHT. 
In section \ref{section:pCltoCl} we
show how an unbiased estimate of the true underlying power spectrum 
can be recovered
from the $\tC_\ell$-s with the aid of the  Monte Carlo simulations.
The tests of the method on simulated  \bldb\  observations are
described in section
\ref{section:montecarlo}, and the application  of the method 
is discussed in section \ref{section:conclusion}.

\section{From Time Ordered Data to Pseudo Power Spectrum}
\label{section:TODtoCl}
Single dish CMB experiments produce for each detector a data stream, 
or the time ordered data (TOD),
of the direction of observation 
and the sky temperature  as
measured through the instrumental beam. 
We  assume that the
beam is known, that it is close to isotropic in the main lobe,
that the side lobes are negligible, and 
and that the pointing at each time is known to an accuracy better
than the size of the main lobe of the beam.
Exceptions to these assumptions will be addressed in section~\ref{section:algo}.
We will also assume that all the TOD samples affected by transient events,
such as cosmic ray hits, have been removed and that in order to
preserve the TOD continuity the resulting gaps are 
filled with fake data having
the same statistical properties as the genuine observations 
(eg, Prunet et al, 2000, Stompor et al, 2000).

\subsection{From TOD to Sky Map}
\label{section:TODtomap}
The data produced by each detector at a time $t$ can be modeled as
\beq
	d_t = P_{tp} \Delta_p + n_t,
	\label{eq:tod}
\eeq
where $\Delta_p$ is the sky temperature, 
that we assume to be pixelised and smoothed with
the instrument beam, $P_{tp}$ is the pointing matrix,
$p$ is the pixel index and $n_t$ is the instrumental noise.

If the TOD noise is  Gaussian distributed  with a known correlation
function $N_{tt'} = \VEV{n(t)n(t')}$, 
the optimal solution for the sky map 
\beq
	m_p = (P^{\dagger}_{pt} N^{-1}_{tt'} P_{t'p'})^{-1} P^{\dagger}_{pt}
N^{-1}_{tt'} d_{t'}
	\label{eq:optimal_map}
\eeq
minimises the residual noise in the pixellised map,
$\Delta_p-m_p$
(Lupton 1993, Wright 1996, Tegmark 1997). While being completely
general
this procedure is impractical for very long TOD streams because of
the required inversion of the large matrix $N_{tt'}$.
A simplification is possible under the assumption of the TOD noise
being piece-wise stationary, and its correlation matrix representable as
circulant, $N_{tt'} = N(t-t')$.
Eq.~(\ref{eq:optimal_map}) can then be solved either directly
(with a computational scaling of $\sim {\npix^3}$), 
or by using  iterative methods as discussed by
Wright (1996), or Natoli et al. (2001)
(in which  case the computation time
is dominated by Fourier space convolutions of the TOD corresponding to the
product $N^{-1}_{tt'} d_{t'}$ in Eq.~\ref{eq:optimal_map}).
Iterative approaches scale like $N_{\rm iter} \ntau \log \ntau$,
where $\ntau$ is
the number of time samples, and $N_{\rm iter}$ is the number of
iterations. $N_{\rm iter}$  depends on the required accuracy of the
final map, and it is of the order of a few tens in the case of a
conjugate gradient method of linear system solution (Natoli et al. 2001).

If the TOD noise properties are  not known beforehand, however, 
as is generally the case, the
Eqs~(\ref{eq:tod}) and (\ref{eq:optimal_map}) can be solved
iteratively together.  This returns at each time
step an estimate of the noise stream, $n(t)$, 
and, hence, of the noise time power spectrum. 
The required computational scaling involves
a somewhat larger $N_{\rm iter}$
(Ferreira and Jaffe 2000, Prunet et al. 2000, Stompor et al. 2000, and 
Dor\'e et al. 2001a).

\begin{figure*}\vbox{
\vspace{-5cm}
\epsscale{2.3}
%{\plotone{plots/combine_map.ps}}
{\plotone{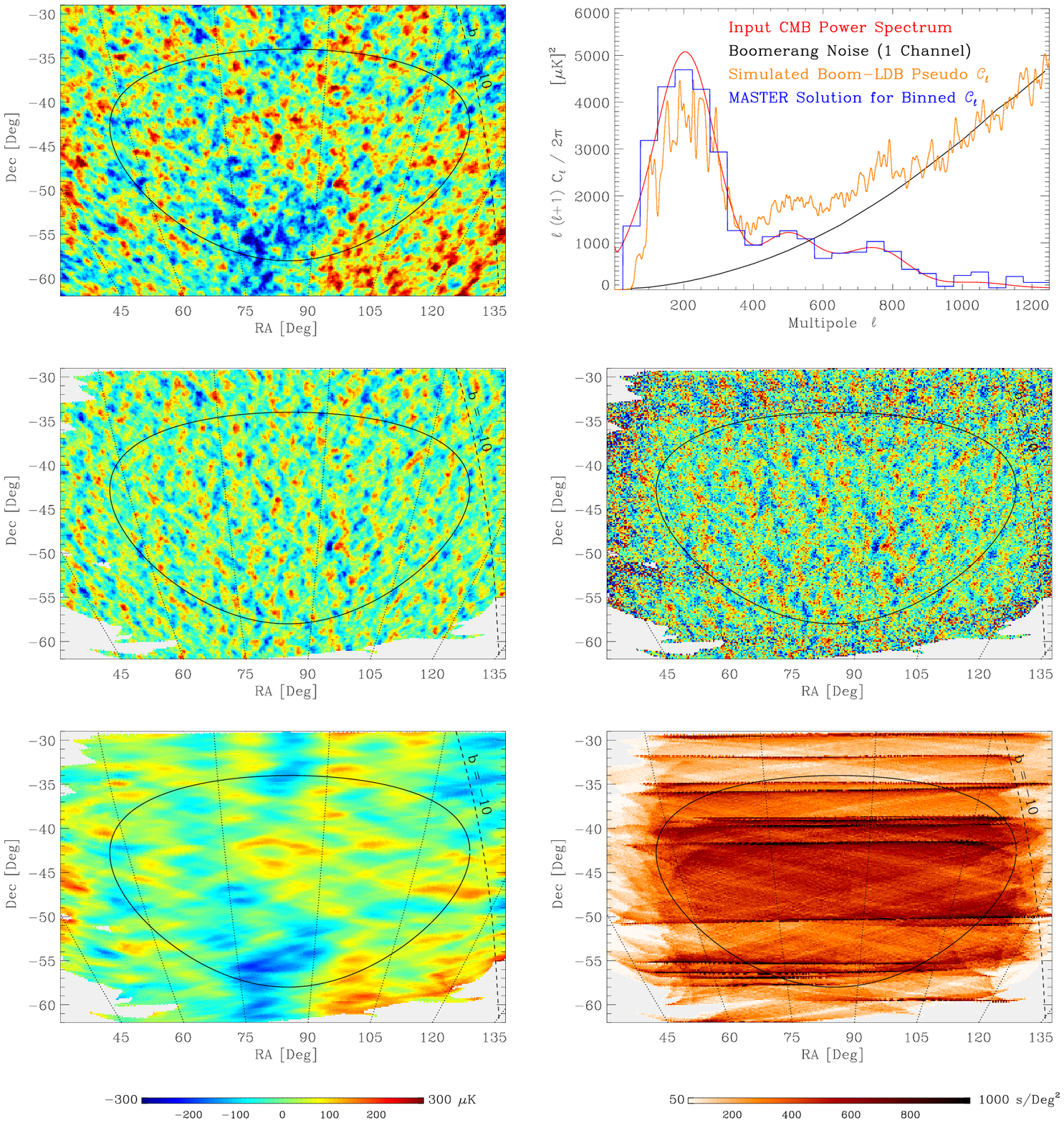}}
\vspace{-4cm}}
\caption{Simulation of the \bldb\ experiment and application of 
MASTER to extract the CMB angular power spectrum.
The oval contour on the maps
shows the ellipse (distorted by  projection) within 
which the power spectrum is estimated ($\fsky=1.8\%$ of the sky).
Top left panel: 
A random realisation of the CMB sky from the theoretical model
described by the power spectrum shown in the top right panel (red
line).
Middle left panel: A noiseless map of the same region of the sky  
made from the TOD with actual \bldb\ pointing and processed with the
100 mHz
high-pass filter (see text).  
Bottom left panel:  the difference between the two CMB sky maps shown
above, which shows the component of the sky signal lost due to 
the combination of Boomerang scanning and data processing.
Middle right panel: A simulation of the
same  Boomerang CMB sky map with  the instrumental noise  included.
Bottom  right panel: Integration time per pixel for 
the actual scanning  of the \bldb\  channel B150A; 
the average integration time is about 500s/Deg$^2$. 
Top right panel: The input power spectrum 
smoothed by the beam and pixel window function (red line); 
The average angular power spectrum
of the instrumental noise (black line);
The pseudo $C_\ell$-s directly measured on the sky map 
shown in the middle right panel and divided by $\fsky$ (orange line);
The binned MASTER estimate of the full sky
power spectrum after removal of noise contribution and correction of
the effect of the high pass filtering and mode-mode coupling (blue histogram).
}
\label{fig:bigone}
\end{figure*}

Since the MASTER method  requires repetitive TOD simulations,
processing,
and map making, the iterative solution of Eq.~(\ref{eq:optimal_map})
can be too time consuming for practical applications.
Therefore, to avoid the necessity to iterate,
we use a suboptimal, fast map making  method
involving the high pass filtering of the TOD stream, which
improves the long time scale behaviour of the noise, 
and reduces the striping of the resulting
map (see section~\ref{section:transfer_function}). 
The map solution is now
\beq
	m_p = (N_{\rm obs}(p))^{-1} \sum_{tt'} P^{\dagger}_{pt} f(t-t') d_{t'},
	\label{eq:naive_map}
\eeq
where $N_{\rm obs}(p) \equiv P^{\dagger}_{pt} P_{tp}$ 
is the number of observations in the pixel $p$,
and $f$ denotes the  high pass filter. 
The computational scaling is now reduced to  $\ntau \log \ntau$.
Clearly, Eq.~(\ref{eq:naive_map}) is only equivalent to
Eq.~(\ref{eq:optimal_map}) if the TOD noise is white, i.e. 
$N_{tt'} = N_0\delta_{\rm Dirac}(t-t')$ (in which case the
filter would be reduced to $f = \delta_{\rm Dirac}(t)$, i.e. no
filtering would be applied).
While the application of the high pass filter reduces  the long term 
noise correlations, it
degrades the CMB signal at low frequencies (see Fig.~\ref{fig:bigone})
and affects the
resulting angular power spectrum derived from the filtered map
solution 
Eq.~(\ref{eq:naive_map}).
This effect  is quantified and corrected for with 
the  Monte Carlo simulations and analysis involving the
filtered map making technique applied to the simulated TODs
of the pure CMB signal.
This procedure will be discussed in detail later on.

\subsection{From Sky Map to Pseudo Power Spectrum}
A scalar field $\Delta T(\n)$ defined over the full sky can be 
decomposed in spherical
harmonic coefficients
\beq
	a_{\ell m} = \int d\n \Delta T(\n) Y^*_{\ell m}(\n),
\eeq
with
\beq
	\Delta T(\n) = \sum_{\ell>0}\sum_{m=-\ell}^{\ell} a_{\ell m} Y_{\ell m}(\n).
\eeq

If the CMB temperature fluctuation $\Delta T$ is assumed to be 
Gaussian distributed,
each $a_{\ell m}$ is an independent Gaussian deviate with
\beq
	\VEV{a_{\ell m}} = 0,
\eeq
and
\beq
	\VEV{a_{\ell m}a^*_{\ell'm'}} = \delta_{\ell\ell'}\delta_{mm'} \VEV{C_\ell},
	\label{eq:ortho_alm}
\eeq
where  $\VEV{C_\ell} \equiv \Cltheory$ is specified  by the  theory of
primordial
perturbations, and parametrised accordingly,
and $\delta$ is the Kronecker symbol.
An unbiased estimator of $\Cltheory$ is given by 
\beq
	C_\ell =
\frac{1}{2\ell+1}\sum_{m=-\ell}^{\ell}\left|{{a}_{\ell m}}\right|^2.
\eeq
$C_\ell$-s are $\chi^2_\nu$-distributed with the mean equal to
$\Cltheory$, 
$\nu=2\ell+1$
degrees of freedom ({\em dof}), and a variance of $2 \Cltheory^2/\nu$.

% Note that Eq.~(\ref{eq:ortho_alm}) implies that the scalar field $\Delta T(\r)$ is
% statistically isotropic.
% If this assumption is true for inflation generated Gaussian fluctuations it is
% not true for most of the foreground contaminations, or 
% if the observation is done with a non isotropic beam, or if the noise filtering
% in the data stream and the scanning strategy induce a an-isotropic reduction of
% signal. {\bf {What the heck ????????????}}

In the case of CMB measurements 
the temperature fluctuations can not be measured
over the full sky, either because of ground obscuration or galactic
contamination for example, and a position dependent weighting $W(\n)$ 
can also be applied to the
measured data, for instance to reduce the edge effects. 
If $\fsky$ represent the sky fraction over which the weighting
applied is non zero, then
\begin{equation}
	\fsky w_i = \frac{1}{4\pi}\int_{4\pi} d\n W^i(\n) 
\end{equation}
is the $i$-th moment of the arbitrary weighting scheme.
The window function can also be expanded in spherical harmonics with the
coefficients  $w_{\ell m} = \int d\n W(\n) Y_{\ell m}^*(\n)$, 
and with a power spectrum
\beq
	{\mathcal{W}}_{\ell} = \frac{1}{2\ell+1}\sum_{m}\left|w_{\ell m}\right|^2,
	\label{eq:power_window}
\eeq
for which 
${\mathcal{W}}(\ell=0) = 4 \pi \fsky^2 w_1^2 $ 
and \\ $\sum_{\ell\ge0} {\mathcal{W}}(\ell) (2\ell+1)
= 4 \pi \fsky w_2$.

%----------------- maps -----------------------------
%----------------------------------------------------

A sky temperature
fluctuation map $\Delta T(\n)$ on which a window $W(\n)$ is applied
can be decomposed in spherical harmonics coefficients
\begin{eqnarray}
	\tilde{a}_{\ell m} &=& \int d\n \Delta T(\n) W(\n) Y^*_{\ell m}(\n)
	\label{eq:alm_cutsky_sum} \\
	& \approx&\Omega_p \sum_p \Delta T(p) W(p) Y^*_{\ell m}(p), 
	\label{eq:alm_cutsky}
\end{eqnarray}
where the integral over the sky is approximated by a discrete sum over
the pixels that make the map, with an individual surface area
$\Omega_p$.

The  pseudo power spectrum $\tC$ can be defined as
\beq
	\tC_\ell =
\frac{1}{2\ell+1}\sum_{m=-\ell}^{\ell}\left|{\tilde{a}_{\ell m}}\right|^2.
	\label{eq:Cl_cutsky}
\eeq

The computation of Eq.~(\ref{eq:alm_cutsky}) for each $(\ell,m)$ up to
 $\ell=\lmax$ 
performed on an arbitrary  pixelisation  of the sphere 
would scale as $\npix\lmax^2$. 
However, if one uses  an adequate lay out of the pixels
to exploit the symmetries of  Spherical Harmonics, such as for example
the ECP (Muciaccia et
al. 1997), 
HEALPix (G\'orski et al. 1998), or Igloo (Crittenden and Turok 1998) 
this computation actually scales like $\npix^{1/2} \lmax^2$. 
In our implementation of the MASTER method,   
after application of the window function on the map, 
the program anafast from the
package HEALPix was used to compute the pseudo power
spectrum.

Wandelt, Hivon \& G\'orski (2000) showed that the marginalised likelihood
$P(\tC_\ell|\Ctheory, N)$ of
the pseudo $\tC_\ell$ for a given underlying theory $\Ctheory$ and 
a given noise covariance $N$ can be computed analytically 
in ${\mathcal{O}}(\npix^{1/2}\lmax^2)$
operations, under conditions of axisymmetric sky window function and
(non necessarily uniform) white noise. Under these assumptions
$P(\tC_\ell|\Ctheory)$ can be used to perform a maximum likelihood fit of
the cosmological parameters to the observed data set. Hansen et al
(2001) extend this approach by using the full pseudo $\tC_\ell$
covariance matrix.
We will now build, starting from the measured $\tC_\ell$, and under more general
conditions on the noise properties and shape of the observing window, 
a new estimator of the full sky power spectrum that can be compared
directly to $\Ctheory$.

\section{From Pseudo Power Spectrum to Full Sky Power Spectrum Estimator}
\label{section:pCltoCl}

The pseudo power spectrum $\tC_\ell$ rendered 
by the direct spherical harmonics transform
of a partial sky map, Eq.~(\ref{eq:Cl_cutsky}), 
is clearly different from the full sky angular spectrum $C_\ell$, 
but their {\em ensemble averages} can be related
by 
\beq
	\VEV{{\tC}_\ell} = \sum_{\ell'} M_{\ell\ell'} \VEV{C_{\ell'}}, 
	\label{eq:model_cutsky}
\eeq
where $M_{\ell\ell'}$ describes the mode-mode coupling resulting from the cut
sky. As described in the appendix, this kernel depends only on the geometry
of the cut sky and can be expressed simply in terms of the power
spectrum 
${\mathcal{W}}_{\ell}$ of the
spatial window applied to the survey (see Eq.~(\ref{eq:kernel_final}) and
Eq.~(\ref{eq:kernel_plane}) for the spherical and planar geometry, 
respectively).

The effect of the instrumental beam, experimental noise, 
and filtering of the TOD stream
can be included as follows
\beq
	\VEV{{\tC}_\ell} = \sum_{\ell'} M_{\ell\ell'} F_{\ell'} B^2_{\ell'} \VEV{C_{\ell'}} +   \VEV{\tN_{\ell}},
	\label{eq:model_rsht}
\eeq
where $B_\ell$ is a window function describing the combined smoothing effects of
the beam and finite pixel size, $\VEV{\tN_{\ell}}$ is the average noise
power spectrum, 
and $F_\ell$ is a transfer function which models the
effect of the filtering applied to the data stream or to the maps. The
determination of each of these terms will be described below.

%The covariance of the pseudo power spectrum is given by ???, which
%could be computed directly in 
%${\mathcal{O}}(\npix^4)$???? operations. We will resort to  Monte Carlo
%simulations to determine these quantities. 

It is often assumed that the 
$\chi^2_{\nu=2\ell+1}$ distribution of $C_\ell$-s on the
full sky can be generalised to cut sky
observations by   scaling $\nu$ to the number of {\em dof} effectively
available. 
Given the large value of $\nu$ the central limit theorem is also invoked
to further simplify this to a Gaussian of the same mean and variance.
From these successive (and excessive) simplifications we will only
retain, 
{\em as a rule of thumb},
that the rms of $C_\ell$ averaged over a range $\Delta \ell$ is approximately
\beq
	\Delta C_\ell \approx \left(C_\ell + \frac{N(\ell)}{B^2(\ell)}\right)
\sqrt{\frac{2}{\nu_\ell}},
	\label{eq:approx_errbar}
\eeq
with 
\beq
	\nu_\ell =(2\ell+1)\Delta \ell\fsky  \frac{w_2^2}{w_4},
	\label{eq:approx_nudof}
\eeq
where the factor $w_2^2/w_4$ accounts for the loss of modes 
induced by the pixel weighting.
We will show 
in section \ref{section:montecarlo} how this compares to the results
of Monte Carlo simulations.

\subsection{Mode-mode Coupling Kernel}
The resolution in $\ell$ 
of the  measured power spectrum is ultimately determined by the
extent of the oberved area of the sky, 
its geometrical shape, and the pixel weighting $W(\n)$ applied to the
survey (see Hobson \& Magueijo 1996, and Tegmark 1996). 
Although we only tested numerically the method on a circular or
elliptically shaped window, 
nothing prevents the use of a more complex window, specially
for a pixel starved experiment with a nonconvex survey area, 
for which  a well designed apodisation could
help improving the achievable spectral resolution. We will show in
section~(\ref{section:montecarlo}) how the choice of window changes
the estimated $C_\ell$ spectrum.

%The computation of kernel $M_{\ell \ell'}$
%requires the calculation of the power spectrum ${\mathcal{W}}$ of the
%window function applied on the sky. If the window $W(\n)$ has some features
%with a scale smaller than the typical 
%inter-pixel distance, such as the sharp edge of a top-hat
%window, using the discrete 
%sum (\ref{eq:alm_cutsky}) instead of the continuous sum
%(\ref{eq:alm_cutsky_sum}) can introduce 
%some artifacts which depends on both the
%pixel size and on the window location. If one wants to include correctly this
%artefact in the analysis,
%it is preferable to compute ${\mathcal{W}}_\ell$ numerically, using the same
%code as the one used to analyse the CMB maps, Eq.~(\ref{eq:alm_cutsky_sum}), 
%with the same settings in term of pixel size and location on the sky. 

\subsection{TOD Filter Transfer Function}
\label{section:transfer_function}
The transfer function $F_{\ell}$ introduced in 
Eq.~(\ref{eq:model_rsht}) describes 
the effect  of any filtering applied to the TOD 
stream or to the map on the angular power spectrum. 
A specific example of the latter is the removal of parallel
stripes extending along a direction different from the scanning direction 
observed in
some channels of the \bldb\  data  (Netterfield et al. 2001, 
Contaldi et al. 2001). 
The
filtering of the TOD has broader applications, however, and can take the
form of a high pass filtering that serves several purposes, as follows:
\begin{itemize}
\item reduce the contribution of the low frequency noise (1/f noise) to the
map, specially if the scanning strategy and/or the map making technique used do
not optimize the removal of these modes,
\item reduce the scan or spin synchronous noise, which may appear at 
the scan frequency and its harmonics,
\item remove from the signal the contribution from the large scale 
anisotropies, which
are poorly constrained on an incomplete sky survey, 
and are likely to contaminate the
estimated power spectrum at all the smaller scales.
\end{itemize}

%==========================================================================
\begin{figure}[t]
\psfig{file=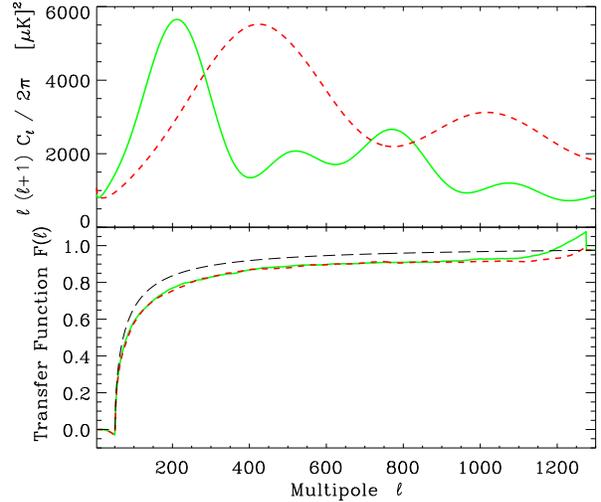,width=0.5\textwidth}
 \caption{Transfer function of the measured angular power spectrum
 corresponding to the sharp high pass filtering at 100 mHz 
 of the  \bldb\ TOD is shown here derived for 
 for two different input CMB anisotropy power spectra. 
 The upper panel shows the two input spectra
 corresponding to a flat Universe (green solid line), 
 and an open Universe (red dashed line). 
 The bottom panel shows the respective transfer functions, which
 differ by less than 1\% up to $\ell=1000$. The black dotted line shows the
 analytical prediction for a toy model of parallel scans
 (Eqs. \ref{eq:approx_tr_bw_low}, \ref{eq:approx_tr_bw_hi}).}
 \label{fig:transfer}
\end{figure}
%==========================================================================

% The determination of the transfer function $F$ introduced in
% Eq~\ref{eq:model_rsht} is obviously complex as this function
% describes in $l$ space an operation performed on the data stream.
It should be noted that the validity of Eq.~(\ref{eq:model_cutsky})
for any sky window relies on the fact that the
statistical properties of the full sky temperature fluctuations are isotropic, as
implied by Eq.~(\ref{eq:ortho_alm}). This assumption is broken by the
high pass filtering of the data stream which creates a preferred
direction on the sky. Therefore the introduction of the single scalar
function $F_{\ell}$ in Eq.~(\ref{eq:model_rsht}) should be
seen as a simplifying ansatz for a more complex reality. We can
numerically test the validity of this ansatz by showing for instance
that $F_{\ell}$ depends weakly on   the choice of the underlying power spectrum.

Given an input CMB power spectrum $\Cltheory$, a number, 
${\nmcs}$, of noise-free full sky
realizations of this spectrum can be simulated (using for example the
program synfast from HEALPix). These maps are then ``observed'' using
the actual scanning strategy, the resulting model TODs projected 
back on the sky using
Eq.~(\ref{eq:naive_map}), and their individual power spectra $\tC_\ell$ are
extracted using Eqs.~(\ref{eq:alm_cutsky}) and (\ref{eq:Cl_cutsky}).

Equation~(\ref{eq:model_rsht}) can then be applied to the
average $\VEV{{\tC}_\ell}_{MC}$ of these measurements to
determine the transfer function. In order to avoid inverting the kernel
$M_{\ell \ell'}$ this system can be solved iteratively. 
In the case of a scanning
experiment, such as Boomerang, Appendix~\ref{append:transfer} shows 
how a model of parallel scan can
render a first order solution $\fzero$. 
In the more general cases, a starting solution
could be $F^{(0)}_\ell = \VEV{{\tC_\ell}}_{MC} / (\fsky w_2 B^2_\ell \Cltheory)$.
If $\VEV{{\tC}_\ell}_{MC}$ is replaced by its running average
$S(\VEV{{\tC}_\ell}_{MC})$ (typically computed over $\Delta \ell = 50$
points) 
it appears that one
iteration is sufficient to obtain a stable estimate of the
transfer function 
\begin{eqnarray}
	F_{\ell} & = &  F^{(0)}_{\ell}  \\ &+ & 
\frac {S\left(\VEV{{\tC}_\ell}_{MC}\right) - \sum_{\ell'} M_{\ell\ell'} F^{(0)}_{\ell'}
B^2_{\ell'} \VEV{C_{\ell'}} }  
{B^2_{\ell}\VEV{C_{\ell}} \fsky w_2 }.\nonumber
	\label{eq:transfer_iter}
\end{eqnarray}

We used Eq.~(\ref{eq:transfer_iter}) to compute the transfer function in Monte
Carlo simulations of \bldb\  observations 
on an elliptically shaped region which  comprises 1.8\% of the sky
(see section \ref{section:montecarlo} for details) 
for two different input power spectra; the first one
corresponding to a flat universe with a first peak 
at $\ell_{\rm peak}\simeq 220$, and the
other one corresponding to an open universe with $\ell_{\rm peak}\simeq
420$. 
Figure~\ref{fig:transfer}
shows that the resulting $F_{\ell}$-s are almost identical, 
with a discrepancy smaller
than 1\% in the range $50<\ell<1000$. This
justifies the use of the simple ansatz~(\ref{eq:model_rsht}) 
as a model of the effect of the
time filtering on the angular power spectrum and demonstrates that the
determination of the transfer function can be done nearly independently of
any assumptions about the actual CMB power spectrum.

Using approximation~(\ref{eq:approx_errbar}) one expects that the
error $\delta F_{\ell}$ done on the MC estimation of the transfer function
decreases as
\begin{equation}
	\frac{\delta F_{\ell}}{F_{\ell}} \approx
\sqrt{\frac{2}{(2\ell+1) 
\Delta \ell f_{\rm sky}}}
	\frac{1}{\sqrt{\nmcs}}.
\end{equation}

So if $\Delta \ell f_{\rm sky} \simeq 1$ and $\ell>100$, 
an estimate of the transfer
function better than 1\% can be obtained in $\nmcs \la 100$ realisations.

The computation of the transfer function $F_\ell$ is required because of the
filtered  map making technique used (Eq.~\ref{eq:naive_map}), which 
alters the signal in 
low frequency modes. 
However, even if a more sophisticated map making were used,
the map obtained is usually not  an unbiased representation of the 
true sky because
of various systematic effects present in the data, 
and the computation of $F_\ell$ is still necessary.

If several detectors with different beams are analysed simultaneously, the 
realisation of the same sky with a different smearing can be used 
as an input to the
TOD simulation of each detector. 
In the analysis of the coadded map an effective beam window
function $B(\ell)$ can be modeled as the weighted average of all individual beams
(see Wu et al. 2001). 
If the actual effective beam were different from this model the
difference would be reflected in the transfer function $F$.

\subsection{Noise Power Spectrum}

From the point of view of modeling the CMB experiment the simplest
form of instrumental noise is the stationary, white, Gaussian process.
Reality however is usually not as simple, the actual experimental
noise is often  non-stationary, ``coloured'', sometimes non-Gaussian,
and correlated to some internal variables of the instrument,
such as its acceleration, the cold
plate temperature, the orientation relative to the sun, to the balloon or
to the ground. For many of those reasons the noise 
often  can not be efficiently averaged out.
However, if
these noises can be modeled to a reasonable accuracy, 
they can be included, at little or no extra cost, in the
Monte-Carlo pipeline described here, and their
effect on the measured $C_\ell$-s can be assessed, and possibly removed.

As mentioned in  \S\ref{section:TODtomap}  an estimate of the noise 
time correlation function $N_{tt'}$ and
its time power spectrum can be extracted from the actual
data stream.
Using this information a fake Gaussian ``noise stream'' can be simulated 
and projected on the sky with the actual scanning strategy using
Eq.~(\ref{eq:naive_map}), with the same high pass filtering $f$. 
The power spectrum $\tN(\ell)$ of the
resulting noise map is extracted according to Eqs.~(\ref{eq:alm_cutsky}), and
(\ref{eq:Cl_cutsky}) and the whole process
is reproduced as many times as necessary to obtain a Monte-Carlo
estimate of the average noise angular power spectrum, $\VEV{\tN(\ell)}_{MC}$.
If several detectors are analysed simulatenously, and their noise
is known to be correlated, these correlations can be included 
in the noise stream simulations.

\subsection{Estimated Power Spectrum}
In order to reduce the correlations of the $C_\ell$-s 
induced by the cut sky, and also to reduce the errors  on 
the resulting power spectrum estimator, it is  convenient to bin the  
power spectrum in
$\ell$. The slowly varying ``flattened'' spectrum
$\mathcal{C}_\ell\equiv \ell(\ell+1)C_\ell/2\pi$ is a preferable
candidate for such  binning.
For a set of $\nbins$ bins, indexed by $b$, 
with respective boundaries $\ell_{\rm low}^{(b)}
< \ell_{\rm high}^{(b)}< \ell_{\rm low}^{(b+1)}$, one can define 
the binning operator $P$ as follows
\begin{eqnarray}
	P_{b\ell} & = & \frac{1}{2\pi}\frac{\ell(\ell+1)}{ \ell_{\rm low}^{(b+1)}- \ell_{\rm
	low}^{(b)} }, \quad 
	{\rm if}\quad 2 \leq \ell_{\rm low}^{(b)} \leq \ell < \ell_{\rm low}^{(b+1)}
	\nonumber \\
	     & = & 0,\quad {\rm otherwise}
\end{eqnarray}
and the binned power spectrum is ${\mathcal{C}}_b = P_{b\ell}C_\ell$.
The reciprocal operator (corresponding to a piece-wise interpolation) then
reads
\begin{eqnarray}
	Q_{\ell b} & = & \frac{2\pi}{\ell(\ell+1)}, \quad 
	{\rm if} \quad 2 \leq \ell_{\rm low}^{(b)} \leq \ell < \ell_{\rm low}^{(b+1)}
	\nonumber \\
	     & = & 0,\quad {\rm otherwise}.
\end{eqnarray}
The two operators above are defined for the flat band, disjoint bins. 
They can be easily modified to
account for an $\ell$ dependent weighting within each bin 
(for example designed to enhance the less noisy multipoles)
without changing the rest of the discussion.

Rewriting Eq.~(\ref{eq:model_rsht}) as
\beq
	\VEV{{\tC}_\ell} = K_{\ell\ell'} \VEV{C_{\ell'}} + \VEV{{\tN}_\ell}
\eeq
we then look for a solution to 
\beq
	P_{b\ell} K_{\ell\ell'} \VEV{C_{\ell'}} = P_{b\ell} \left( \VEV{{\tC}_\ell} -
\VEV{{\tN}_\ell} \right).
\eeq

This system has $\lmax$ unknowns for $\nbins$ equations. 
We seek the solution such that  
$\VEV{\mathcal{C}_\ell} = \ell(\ell+1)\VEV{C_\ell}/2\pi$ 
is a piece-wise constant. 
If we
replace $\VEV{C_{\ell}}$ by $Q_{\ell b} P_{bl'} \VEV{C_{\ell'}} 
				 =  Q_{\ell b} \VEV{\mathcal{C}_{b}} $
then
\beq
	\VEV{\mathcal{C}_{b}} = K_{bb'}^{-1} P_{b'\ell} \left( \VEV{{\tC}_\ell} -
\VEV{{\tN}_\ell} \right),
	\label{eq:c_bin}
\eeq
where 
\begin{eqnarray}
	K_{bb'} &=& P_{b\ell} K_{\ell\ell'} Q_{\ell'b'}, \nonumber \\
		&=& P_{b\ell} M_{\ell\ell'} F_{\ell'} B^2_{\ell'}
		Q_{\ell'b'}.
	\label{eq:binkernel}
\end{eqnarray}
%%%%%------- binned kernel ------------------------
\begin{figure*}[ht]
\epsscale{2.35}
%{\plottwo{plots/kern_full.ps}{plots/kerninv_full.ps}}
%{\plottwo{plots/kern_cross.ps}{plots/kerninv_cross.ps}}
{\plottwo{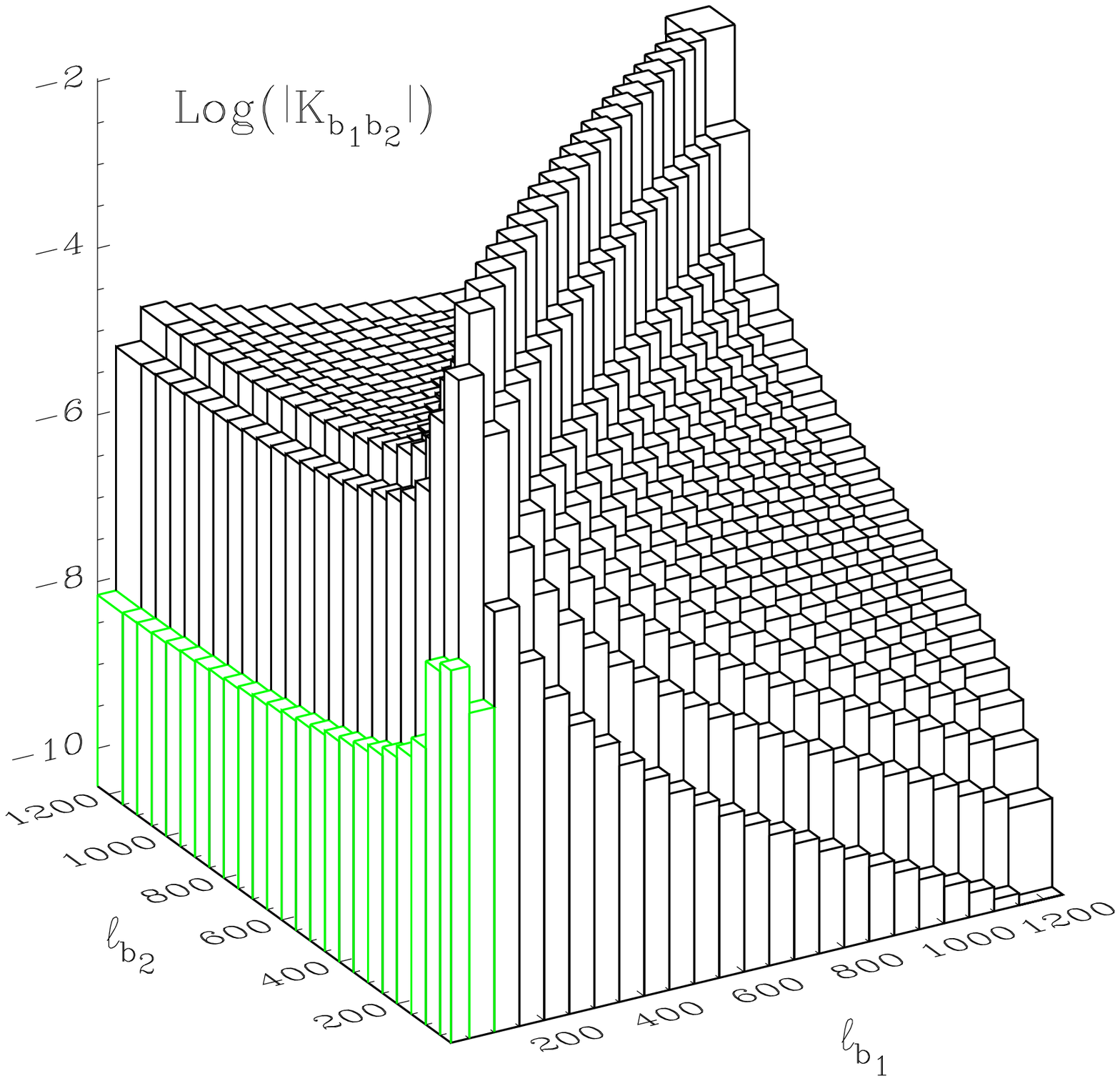}{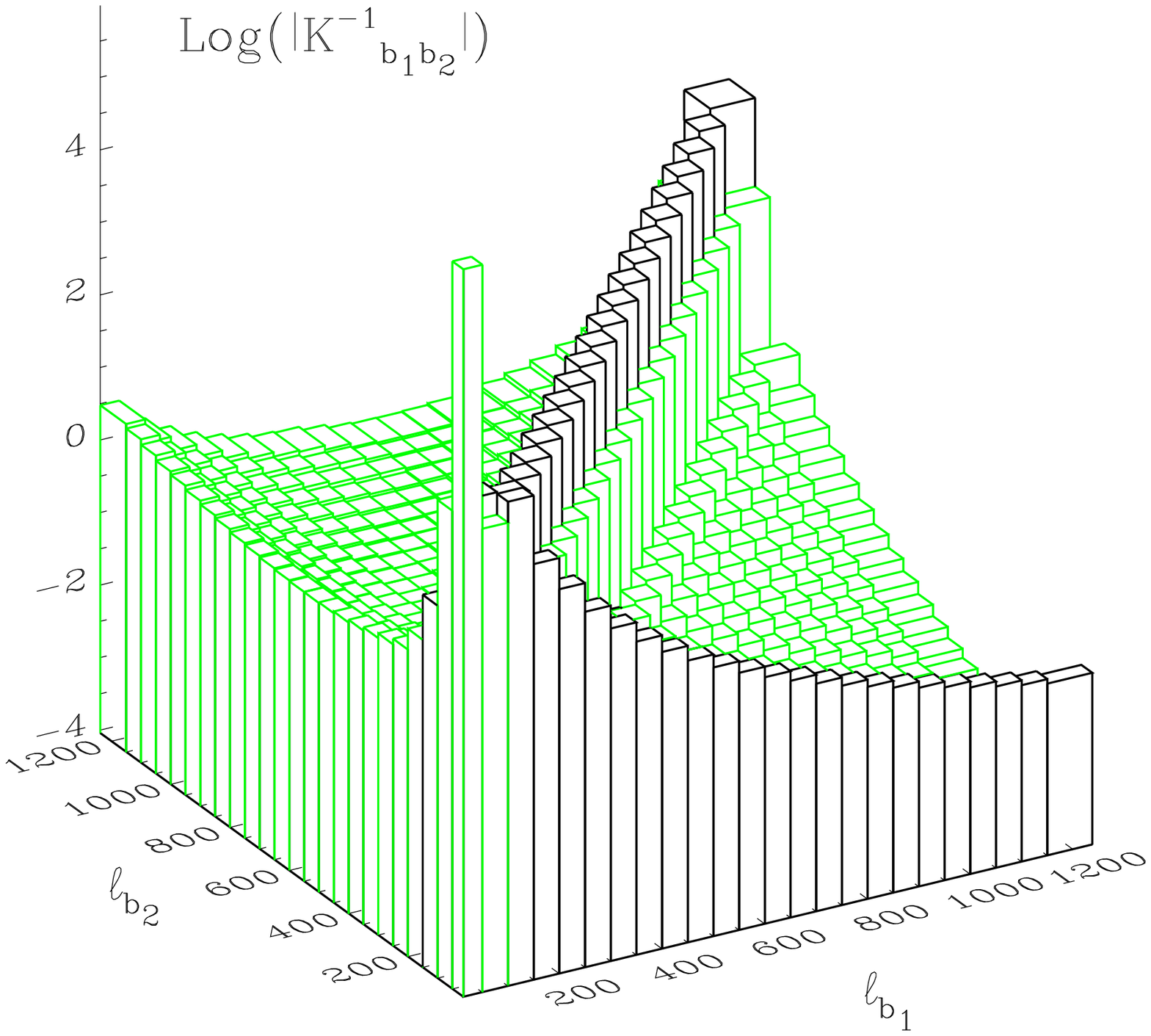}}
{\plottwo{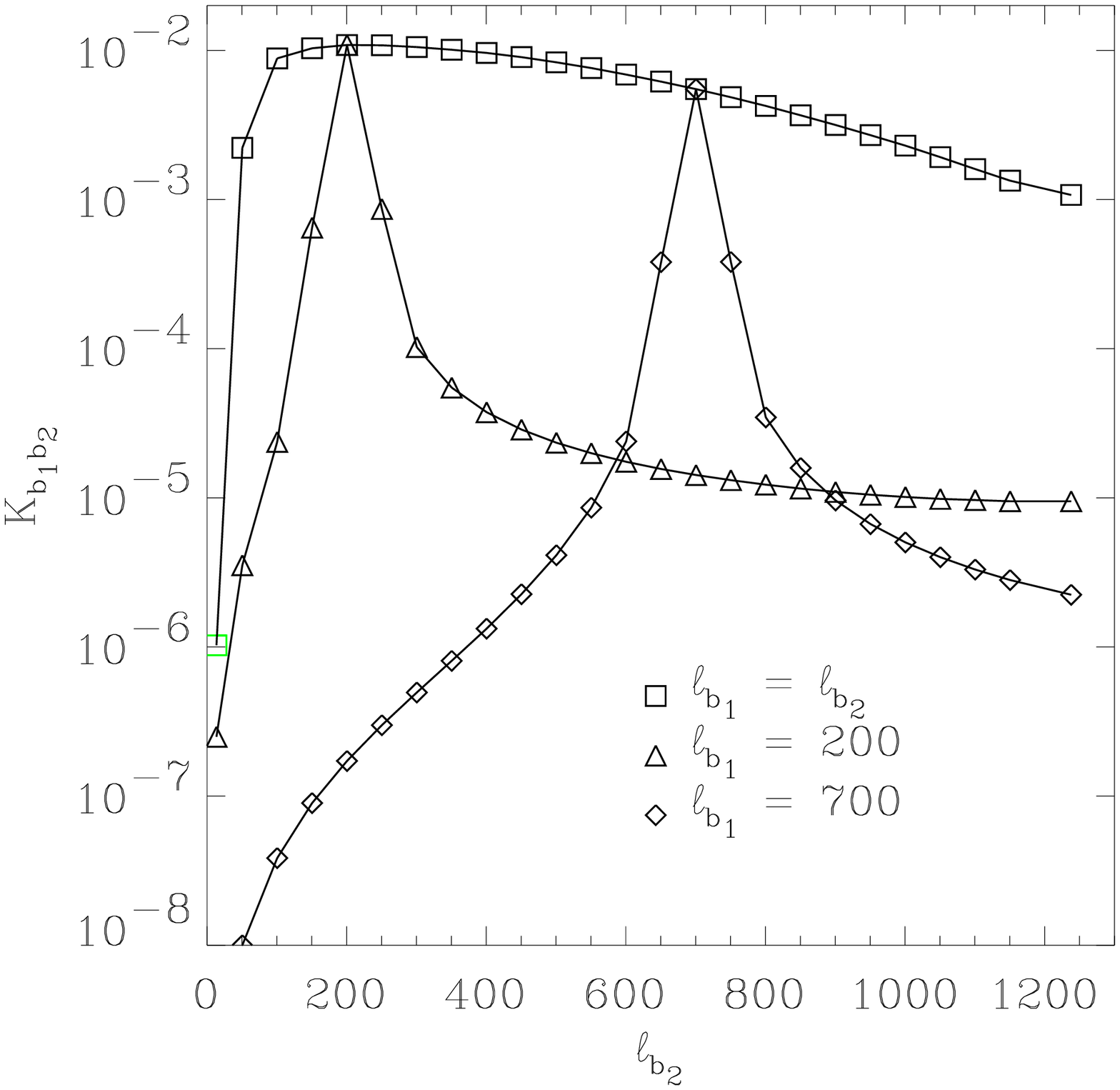}{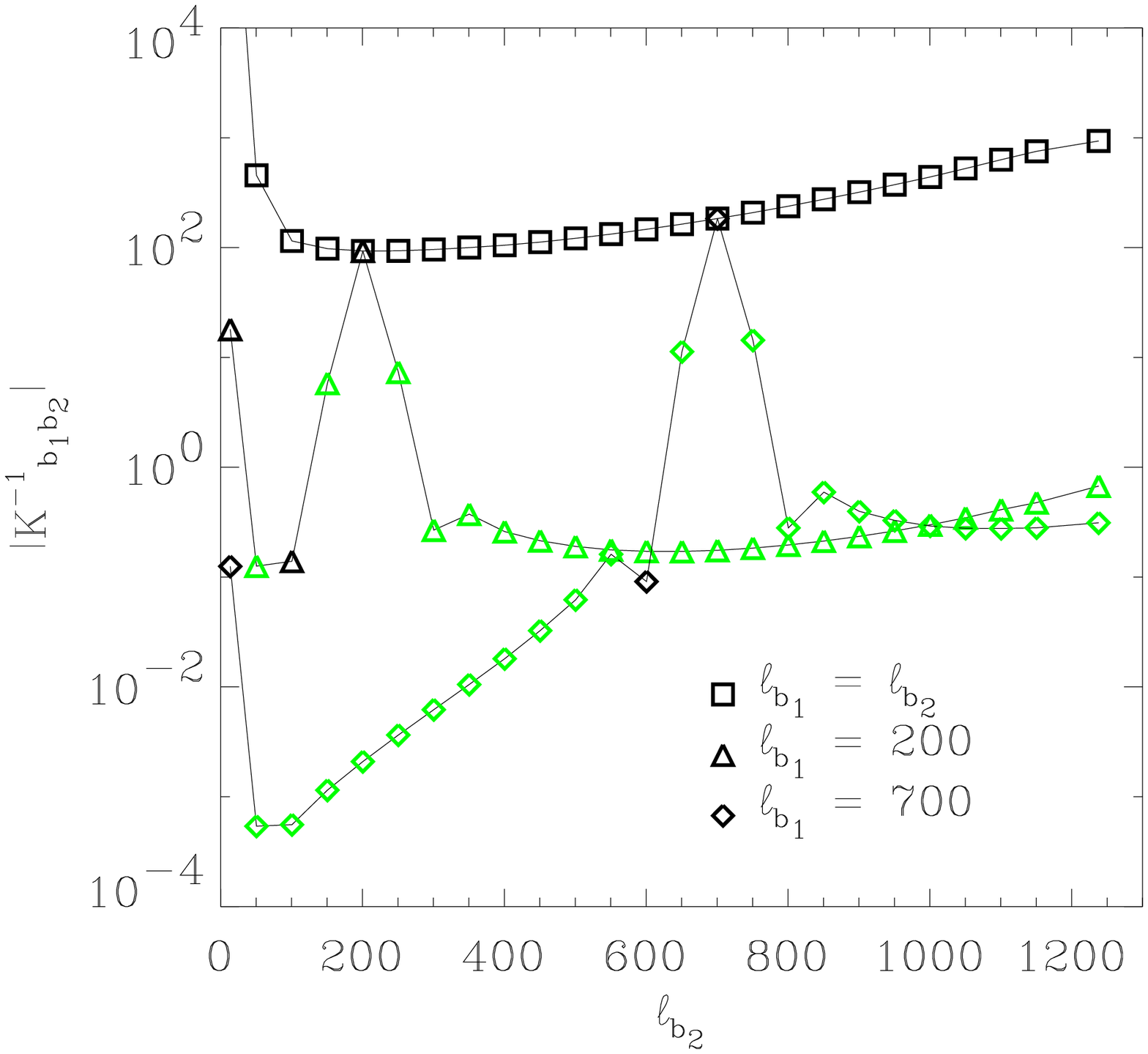}}
\label{fig:kernel}
\caption{Binned power spectrum coupling kernel $K_{b b'}$ and its inverse
(absolute values shown, with green color indicating the negative elements)
for an elliptically shaped top hat sky window
covering 1.8\% of the sky.
Binwidth is $\Delta\ell=50$, except for the last bin, for which 
$\Delta\ell=150$.
The diagonal elements and the $\ell_{b_1} = 200,\, {\rm and}\, 700$
rows of both matrices are shown in the bottom panels. 
}
\end{figure*}
%%%%%-------------------------------

An unbiased estimator ${\widehat{\mathcal{C}}}_b$ of the whole sky power
spectrum is then given by
\begin{equation}
	\widehat{\mathcal{C}}_b = K_{bb'}^{-1} P_{b'\ell} \left( {\tC}_\ell -
\VEV{{\tN}_\ell}_{\rm MC} \right).
\label{eq:Cb_estim}
\end{equation}
An estimator of the noise ``on the sky'' can also be introduced
\begin{equation}
	\widehat{\mathcal{N}}_b = K_{bb'}^{-1} P_{b'\ell} 
\VEV{{\tN}_\ell}_{\rm MC} .
\label{eq:Nb_estim}
\end{equation}

Figure~\ref{fig:kernel} shows the kernel $K_{bb'}$ and its inverse for
the configuration described in section~\ref{section:montecarlo}.

It should be noted that the binning of the $\ell$ 
space is performed at the last stage of the
analysis and can be chosen after the MC simulations are done.

\subsection{Covariance Matrix of the Estimated Power Spectrum}
In order to be able to extract the cosmological parameters 
from the power spectrum
estimated as described above, 
one needs to know the errors on each ${\mathcal{C}}_b$, and the
correlations between the bins. This information is contained in the 
covariance  matrix of the estimated power spectrum, which can be
estimated as follows.
A smooth interpolation of $\widehat{\mathcal{C}}_b$ is used  
as the underlying CMB power spectrum, and
a new set of Monte Carlo simulations, including both signal 
and noise, as well as all
the experiment peculiarities, is generated, analysed, and 
reduced in the same way
as the real data. This generates a set of binned power
spectrum estimators $\{{\widehat{\mathcal{C}}}_b\}$.
The elements of the correlation matrix are defined as
\beq
	{\bf C}_{bb'} = \left\langle\left({\widehat{\mathcal{C}}}_b -\VEV{{\widehat{\mathcal{C}}}_b}_{MC}\right) \left({\widehat{\mathcal{C}}}_{b'}
-\VEV{{\widehat{\mathcal{C}}}_{b'}}_{MC}\right)\right\rangle_{MC}.
\label{eq:cov_mat}
\eeq

The error bars on ${\widehat{\mathcal{C}}}_b$ are then given by the square root of the
diagonal elements of ${\bf C}$
\beq
	\Delta {\widehat{\mathcal{C}}}_b = {\bf C}_{bb}^{1/2}.
	\label{eq:mcsht_error}
\eeq

% From Eq.~(\ref{eq:approx_errbar}) one expects the uncertainty on the error bar
% $\Delta(\Delta C_b)$ to decrease like $\approx 1/\sqrt{\nmc}$

\subsection{The Algorithm}
\label{section:algo}
Assuming that for a given CMB experiment the instrumental beam and 
the time domain power spectrum of the instrumental noise
are known, the process of estimation of the full sky power spectrum
from the noisy observations of CMB temperature fluctuations
can be summarised as follows.

\noindent The required tools are :
\begin{itemize}
\item[(a)] a simulation facility for generation of the random
realisations of the CMB sky (e.g.  synfast from HEALPix),
\item[(b1)] a software model of the  experiment which simulates
observations of the sky using the appropriate scanning strategy (Eq.~\ref{eq:tod}), and
generates the model CMB  signal TOD streams,
\item[(b2)] a noise simulator that can generate random realisations 
of the noise with an appropriate power spectrum; possible 
non-gaussian noise features or cross-correlations between detectors 
should  be added at this stage,
\item[(b3)] a fast map making facility, which implements 
Eq.~(\ref{eq:naive_map}), 
and accounts for any
alterations of  the observed TOD stream and/or the produced map,
\item[(c)] a software to compute the pseudo power spectrum 
(Eqs. \ref{eq:alm_cutsky}, and
\ref{eq:Cl_cutsky}) from a given apodised cut sky map (e.g.  anafast
from HEALPix).
\end{itemize}

After the choice of the sky window apodisation function is made 
the procedure
of estimation of the power spectrum involves the following steps:

\begin{enumerate}
\item  Eq.~(\ref{eq:kernel_final}) is used to evaluate the $C_\ell$
coupling
kernel $M_{\ell\ell'}$, which accounts for the effects of limited sky
coverage and apodisation;
\item  a number $\nmcs$ of noise free
Monte Carlo simulations of the observed TOD (produced with (a) and
(b1),  projected on the sky with (b3), and analysed with (c))
are used to estimate the transfer function $F_{\ell}$ 
of any filtering that is applied  to the
actual TOD stream;
\item a number $\nmcn$ of pure noise Monte Carlo simulations of the TODs 
(made with (b1) and (b2), and then projected and
analysed with (b3) and (c))
are used to estimate the angular
power spectrum $\VEV{\tN}_{MC}$ of the noise projected on
the sky;
\item the experimental TOD is coverted into a map using (b3) and its
pseudo power spectrum ${\tC}_\ell$ is obtained with (c);
\item next a set of $\ell$-bins is defined, and 
an estimate of the underlying full sky {\em binned} power spectrum is 
computed using Eq.~(\ref{eq:Cb_estim});
according to Eq.~(\ref{eq:c_bin})  this is an unbiased estimator ---
$\VEV{{\widehat{\mathcal{C}}_b}} = \VEV{{{\mathcal{C}}_b}}$;
this will be demonstrated in section~\ref{section:montecarlo} with simulated
Boomerang observations in which the input spectrum is known;
%\item the error bars on the binned power spectrum are obtained by
%computing from the ensemble of $\nmcsn$ simulations 
%the covariance matrix ${\bf C}_{b b'}$ 
%(Eq. \ref{eq:cov_mat})
%and using its 
%diagonal elements (Eq.\ref{eq:mcsht_error}).
\item the covariance matrix ${\bf C}_{b b'}$ (Eq. \ref{eq:cov_mat}) is computed from $\nmcsn$
simulations of the whole experiment, and the error bars on the binned
power spectrum are obtained from its diagonal elements (Eq.\ref{eq:mcsht_error}).
\end{enumerate}

We assumed the physical beam to be close enough to axisymmetric so that its
smoothing effect on the temperature map is independent of the payload
attitude along the line of sight. It it were not the case, a direct
integration of the temperature over the beam would have to be
performed for each time sample in Eq.~(\ref{eq:tod}).
This operation could be very intensive for extremely structured beams, unless
some symmetries in the scanning strategy, such as the one expected for
satellite missions, allow for a fast convolution implementation (Wandelt \&
G\'orski 2000, Challinor et al, 2000)

On the other hand, the effect of a pointing inacurracy, whether it is
axisymmetric or not, can easily be included in the method by modifying
Eq.~(\ref{eq:tod}) accordingly.

%-------------------------- noise ---------------------------------------
\begin{figure}[t]
\psfig{file=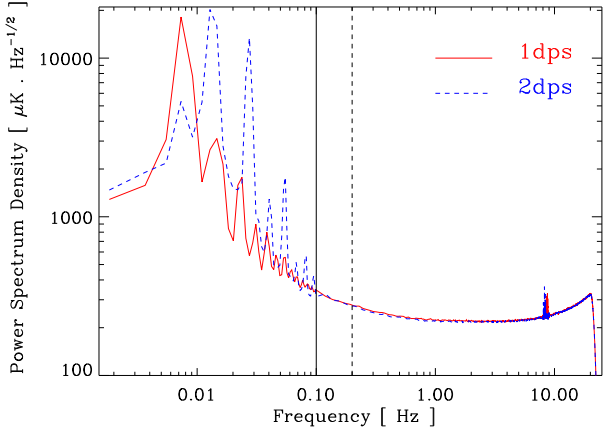,width=0.5\textwidth,angle=90}
\caption{Spectral densities of the noise measured for one of the \bldb\ 
channels is shown for the two scanning speeds (1 and 2 degrees per second) 
of the actual CMB observations. These spectra were 
used to generate the simulated  observations. A sharp high-pass
filter
was applied to the measured and simulated TODs
at 100 and 200 mHz, respectively.}
\label{fig:noise}
\end{figure}
%----------------------------------------------------------------------

\subsection{Computational Scaling}
The algorithm is
maximally parallelisable as each Monte Carlo cycle of the algorithm 
can be performed on a 
separate CPU. 
Therefore the total time required for completion of the estimation of
the power spectrum
from the single detector  CMB observations, with 
${\nmc}={\nmcs}+{\nmcn}+{\nmcsn}$ cycles run on  ${n_{\rm CPU}}$ processors, is given by
\beq
	T_{\rm ~total} = \frac{\nmc}{n_{\rm CPU}} T_{\rm ~MC~cycle}
\eeq
with
	$T_{\rm ~MC~cycle} = T_{(a)} + T_{(b)} + T_{(c)}$, where 
(a),(b), and (c) refer to the CPU time consumption by the 
simulation tools described in section~\ref{section:algo} 
(CMB map
synthesis,  observation simulation and map making, 
and map analysis, respectively). 
In the case of joint multi-detector
analysis the  stage (b) has to be repeated for
each detector, 
whereas the stage (a) only has to be repeated  for each different beam.
In our own implementation (see 
section~\ref{section:montecarlo} for detailed specifications)
the  MASTER method is executed on personal
computers (PCs) equipped
with 850 MHz AMD Athlon CPUs, the following performance is achieved
(on each processor):
\beq 
	T_{(a)} = T_{(c)} \simeq 300\  {\rm s}\  
\left(\frac{\npix}{3\ 10^6}\right)^{1/2} \left(\frac{\lmax}{1300}\right)^2,
\eeq
and 
\beq 
	T_{(b)} \simeq 300\  {\rm s}\  \left(\frac{\ntau}{5\
10^7}\right) 
\left(\frac{\log \nfft}{\log (5\ 10^5)}\right),
\eeq
where $\npix$ is the number of pixels over the whole sky at the chosen
map
resolution (not in the cut sky map), $\ntau$ is the number of time
samples in the TOD set used,
and $\nfft$ is the
typical number of time samples on which the required FFTs are computed.
This leads  to an overall CPU time requirement of
\beq
	T_{\rm total} = 0.5\ {\rm day}\ 
\left(\frac{\nmc}{300}\right)\left(\frac{8}{n_{\rm CPU}}\right).
\eeq
%--------------- figure : montecarlo test --------------------
\begin{figure*}[ht]
\vbox{
\epsscale{2.2}
%{\plottwo{plots/errbar_th_50_cl.ps}{plots/errbar_th_50_mat.ps}}  
%{\plottwo{plots/errbar_c_50_cl.ps}{plots/errbar_c_50_mat.ps}}  
%{\plottwo{plots/errbar_g_50_cl.ps}{plots/errbar_g_50_mat.ps}}
{\plottwo{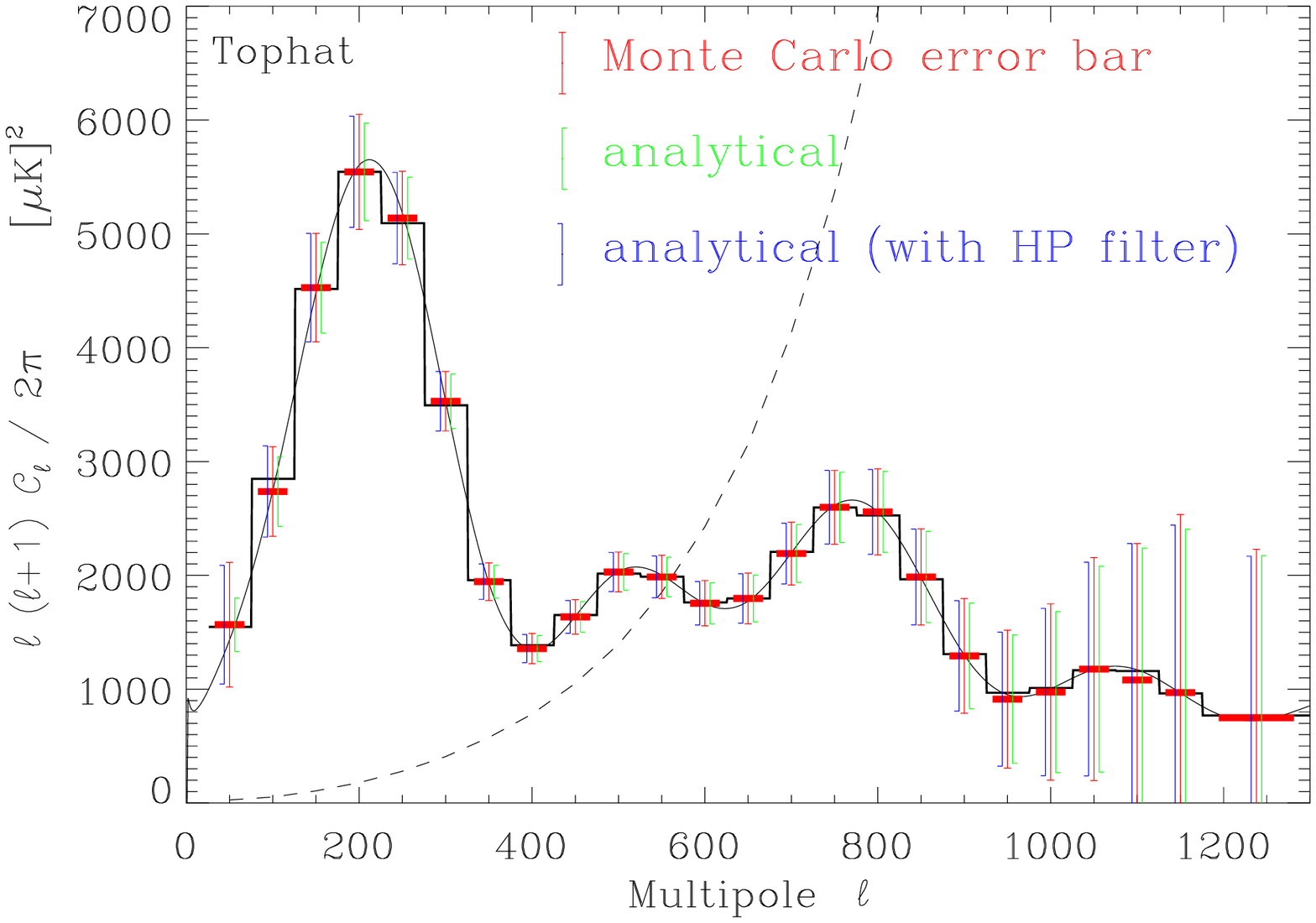}{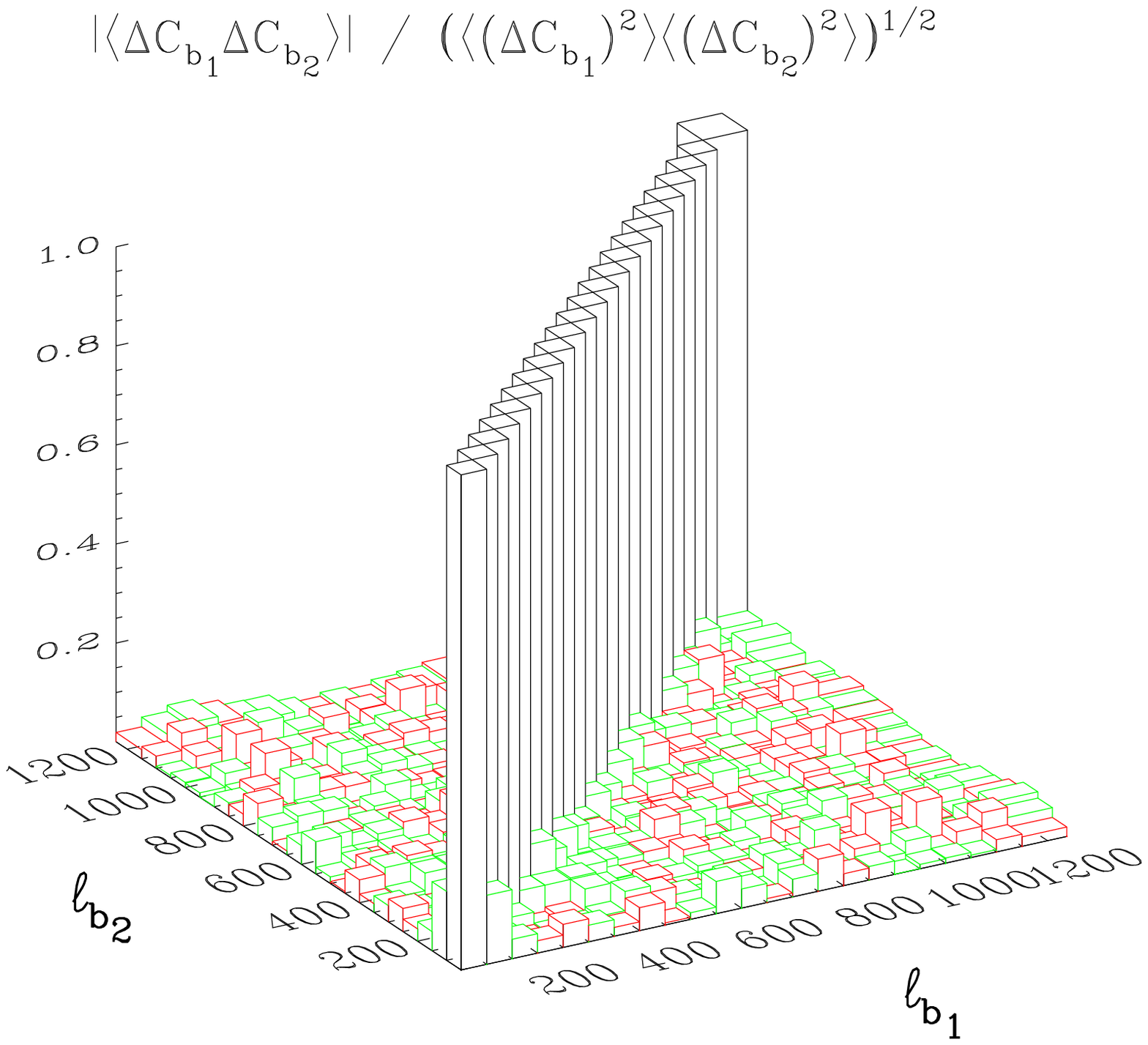}}
	  
{\plottwo{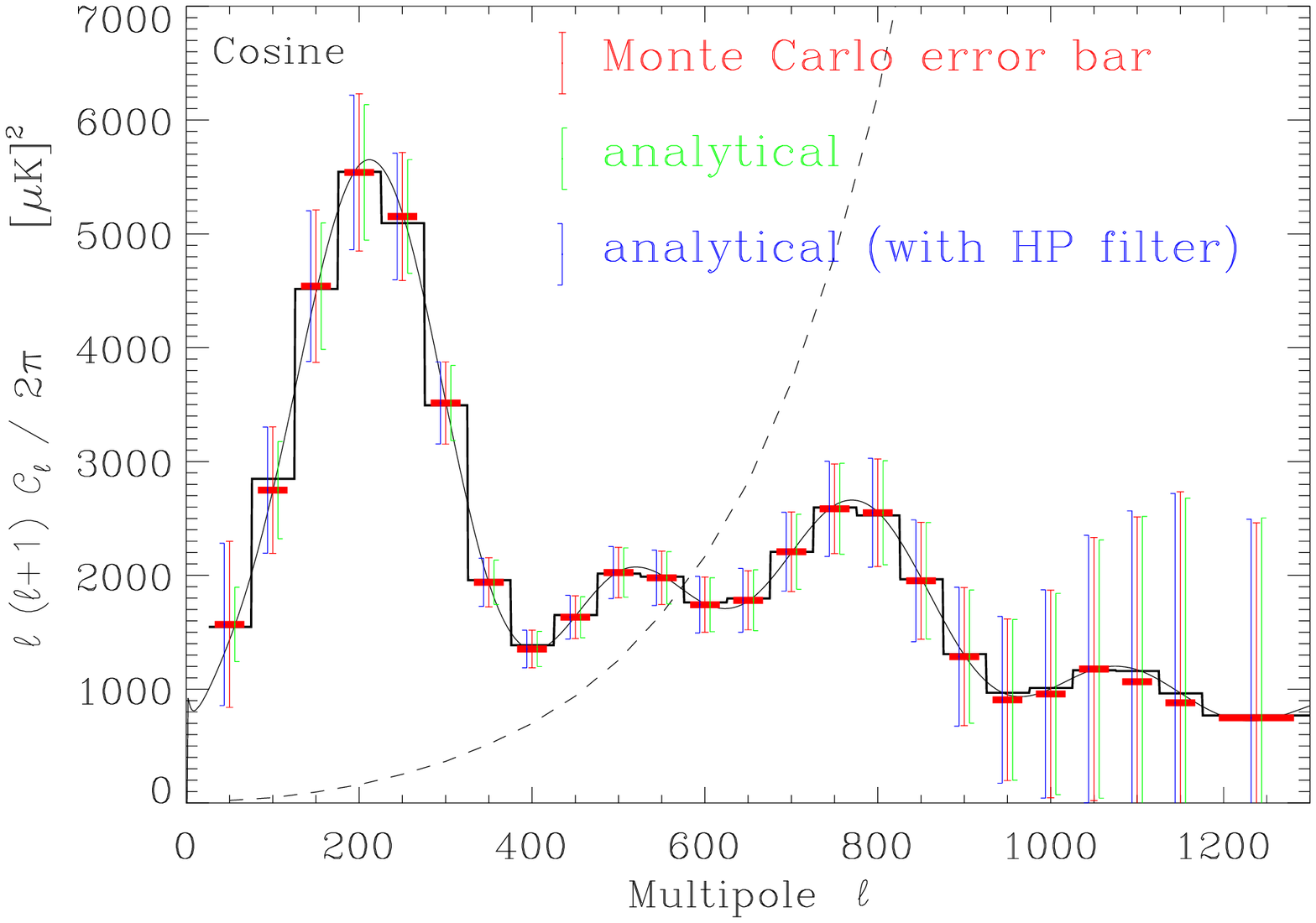}{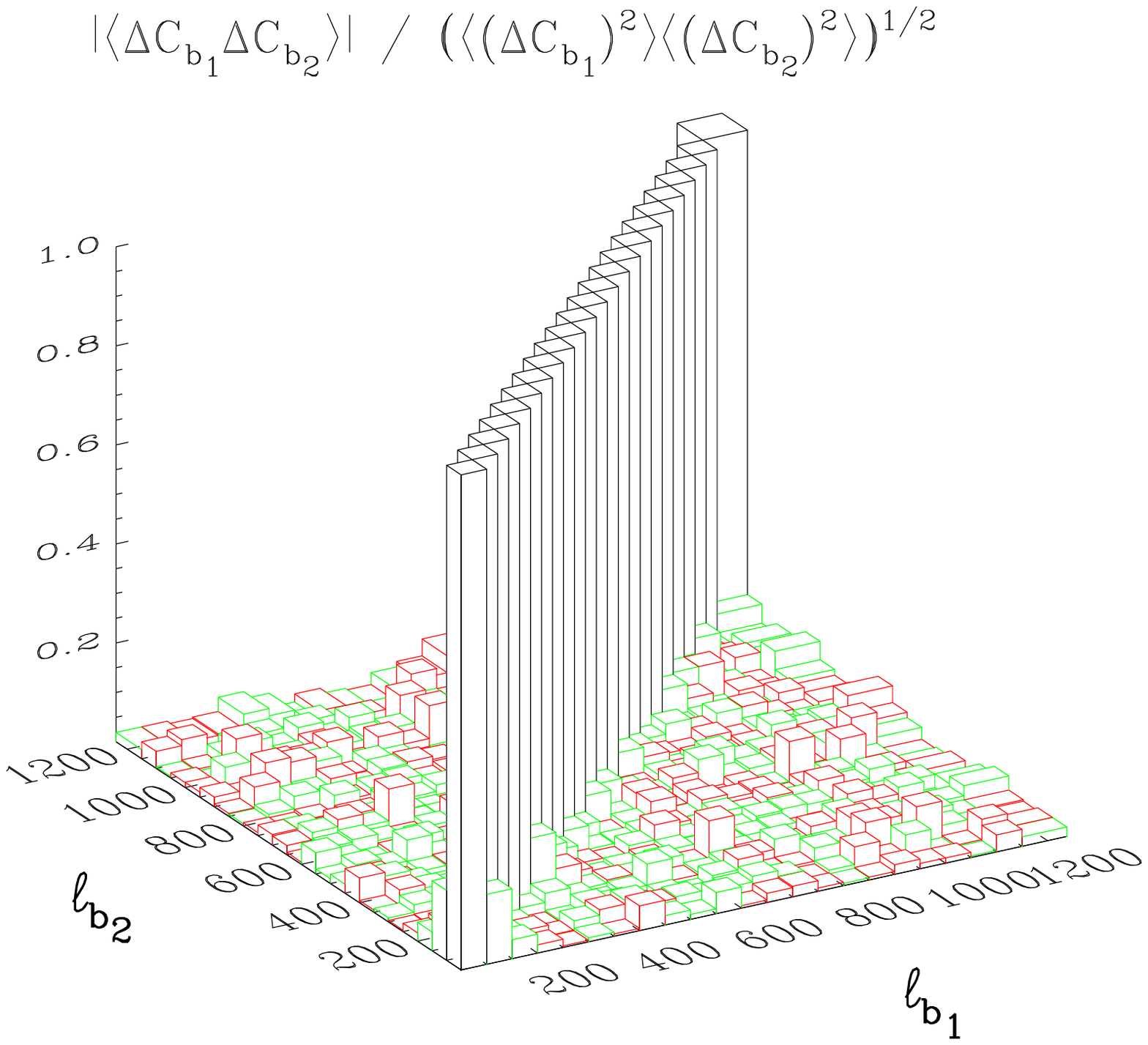}}
	  
{\plottwo{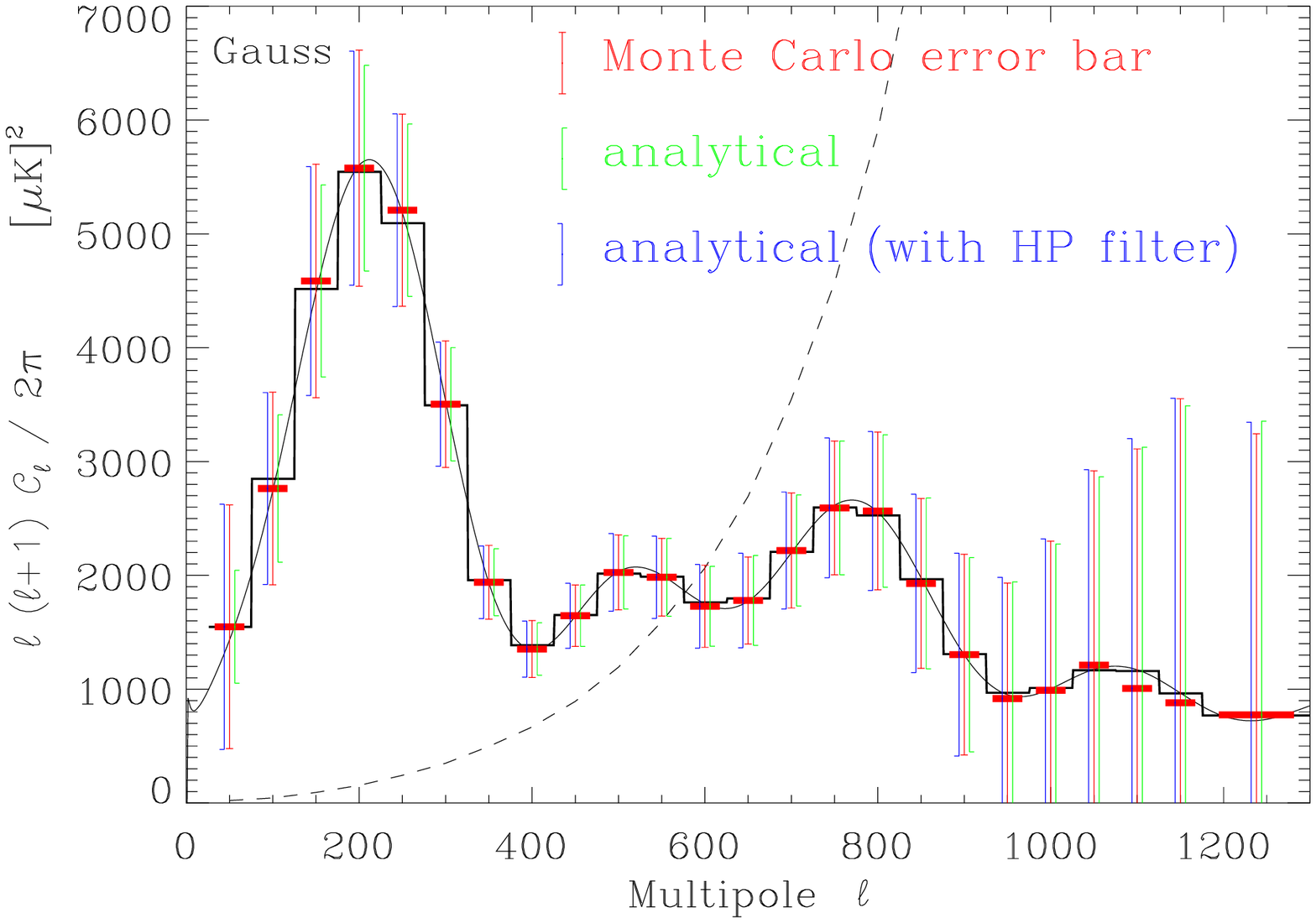}{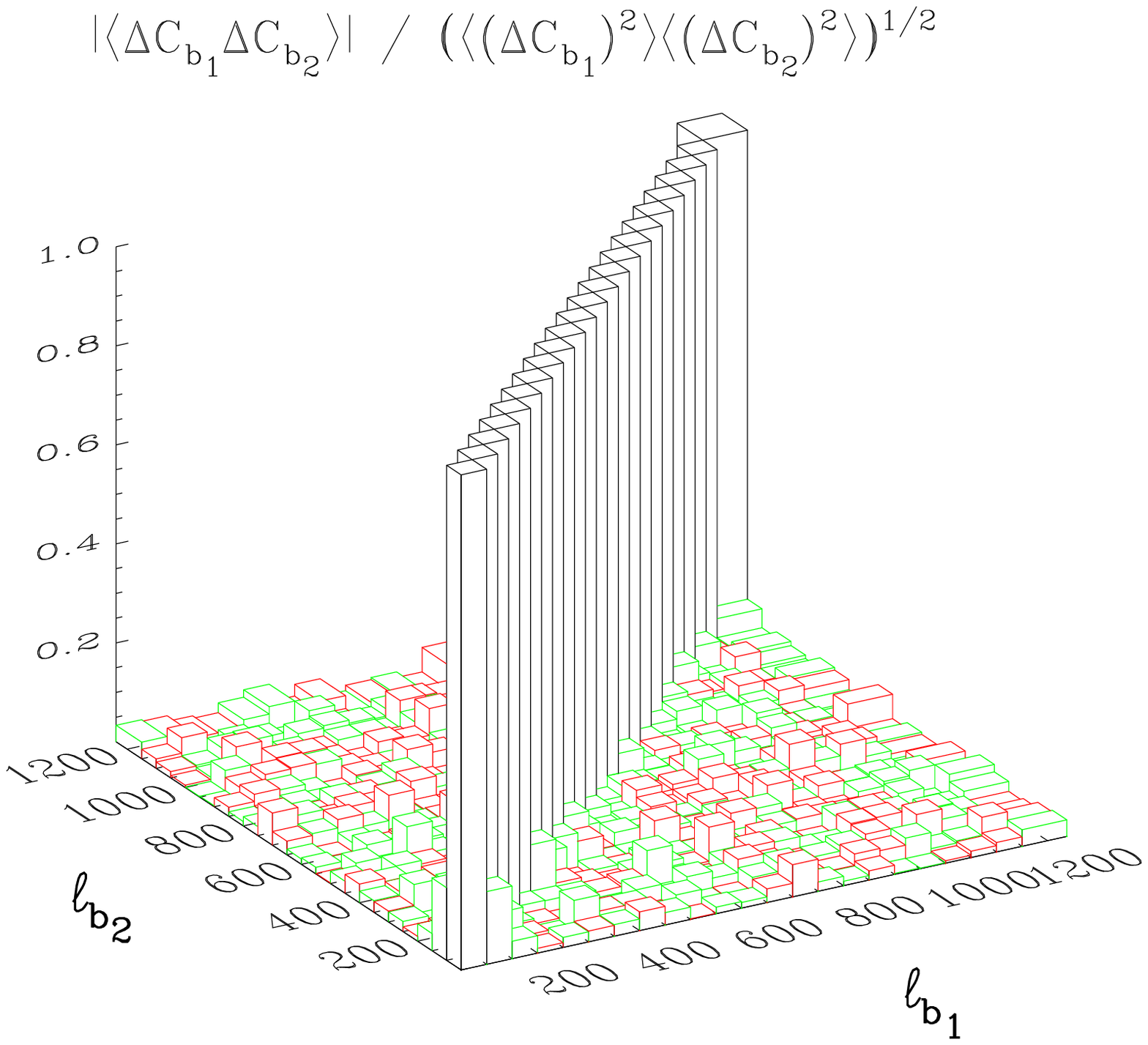}}

}
\caption{Results of the test of accuracy of our method
of estimation of the CMB power spectrum  are
shown in application to the  simulated  \bldb\  data (with known input power
spectrum) with three different sky window apodisations (top hat, cosine
and Gaussian from top to bottom panels). We used an ensemble of 1352
Monte Carlo simulations.
Left panels:  
The input power spectrum is shown as a 
solid black line, and the black histogram shows its bin-averaged
values. The dashed lines show the average spectra of the simulated 
noise.
The red histogram shows our MASTER ensemble 
mean values of the estimated bin-averaged power spectra, which 
are in perfect agreement (to an error on the mean) 
with the input bin-averaged theory.
For each spectral bin three error bars are shown, all centered vertically 
on the mean values of the MASTER estimates,
$\langle{\widehat{\mathcal{C}}_b}\rangle_{MC}$, 
and, for clarity, spread
around
the $\ell$-centers of the bins.
The central (red) error bars show the rms values of the MASTER
ensemble of estimates, while
the right-, and left-shifted (green and blue, respectively) ones
show  the theoretical error estimates based on Eqs. (\ref{eq:approx_nutrans},
\ref{eq:optim_error}) derived, respectively, without and with an
inclusion of the effect of the TOD high-pass filter related spectral
transfer
function $F_{\ell}$.
Right panels: absolute value of normalized binned spectrum correlation matrix 
${\bf C}_{bb'}/\left({\bf C}_{bb}{\bf C}_{b'b'}\right)^{1/2}$ 
(see text). Elements with value in $\left[-0.2,0\right]$ are
represented in green, those with value in $\left[0,0.2\right]$ in red
and those larger than 0.2 in black.}
\label{fig:test_mc}
\end{figure*}
%-------------------------------------------

These timing estimates  could
be easily improved if some repetitive tasks, such as 
the translation from 
sky coordinates to pixel index in (b), or the evaluation of Legendre
polynomials in (a) and (c) are precomputed and stored on disc.

NB: The most daunting challenge  currently foreseen  for CMB anisotropy
analysis  arises
in  the process of reduction and interpretation of the data
that will be collected during the ESA Planck mission (currently
scheduled for launch in 2007). Let us recompute our CPU time
requirements to match the parameters which describe some of the Planck
High Frequency Instrument specifications.
One channel of the HFI with a beam resolution of 5 arcmin should,
during one year of observations,
return ${\ntau} \simeq 5.6\times 10^9$ TOD points, which will be
used to make a HEALPix full sky map with $\npix = 5\times 10^7$ pixels
(of average size of 1.7 arcmin) and which in turn should allow us to estimate 
the angular power spectrum of the sky signals up to $\lmax = 3000$.
With these specifications, and assuming $\nfft\simeq 1.5\times 10^7$ (one day), the CPU time required for execution of our current
implementation of MASTER becomes

\beq
	T_{\rm ~total} \simeq 24\ {\rm days}\ 
\left(\frac{\nmc}{300}\right)\left(\frac{8}{n_{\rm CPU}}\right)
\eeq
on currently available PCs. 
Since dedicated computer servers with 
32, or more, processors at least twice
as fast as  the PCs that we used  
are already presently available, the quoted total execution time for
the MASTER method can be reduced to about 3 days. 
The extra speed up of the CPUs  between now and
the launch of Planck (a factor of $\sim 16$ according to Moore's law)
should enable the
simultaneous analysis of several Planck channels, 
with more sophisticated map making and a larger number of MC cycles 
to improve the power spectrum estimation accuracy, 
in a total time  of a few days.
This means that the MASTER approach is 
fully practical from the point of view of demands related to
scientific analysis of the Planck data.

\section{MASTER Tests on Simulated  Boomerang Observations}
\label{section:montecarlo}
\subsection{Simulations of CMB Observations During The Long Duration
Baloon Flight of Boomerang Experiment}

The MASTER method was tested for application
to the extraction of the CMB anisotropy power spectrum from
the data collected by the {\bldb}\  experiment.
These CMB observations comprise a total of  about $50\ 10^6$ samples
which cover about 4.4\% of the sky, and were acquired at two
different azimuthal sky scan rates of 1 and 2 deg per second 
(for more details on the {\bldb}   flight see Crill, 2001).

We have modeled these CMB  observations of the simulated CMB sky with 
a scanning pattern identical to that of one of {\bldb}   channels. 
The instrumental noise generated in this channel 
was assumed to be Gaussian with a time power
spectrum identical to the one measured during the observations.
The characteristic features of this noise power spectrum
include  a $\sim 1/f$
behavior at low frequency, a knee-frequency of about 100mHz, a
white noise level of 130 $\mu$K.s$^{1/2}$, a series of lines
located at
the harmonics of the scanning frequency (8 mHz at 1dps, and 16 mHz at
2dps), and some microphonic artifacts 
at 8 Hz (see fig~\ref{fig:noise}). 
Angular resolution of this channel was  FWHM $\approx 10$ arcmin, and
we used the actual measured beam profile in the calculations.

The high-pass filtering applied during the process of  map-making
(Eq.\ref{eq:naive_map}) was a sharp
cut off at 100 mHz for the 1dps scan rate, and 200 mHz for the  2dps
scan rate.

The sky maps where pixelised using HEALPix with 7 arcmin pixels
($N_{\rm side} = 512$).
The power spectrum was computed from a subset of the data
on an elliptically shaped region of semi-axes $a=20$ and $b=12$ deg,
which covers
$\fsky = 1.82\%$ of the sky and is centered on the 
best  observed region of the \bldb\  flight. 
As shown in Fig.~\ref{fig:bigone} the sky coverage in this area, 
and therefore
the noise per pixel, is nonuniform.
The number of observations  per pixel varies between 50 and 1510,
with an average of 370, and a standard deviation of 120.
The CMB dipole was not included in the sky simulations 
as the high pass filtering used in map making
reduces its rms residual variation in the observed region of the sky 
to less than 0.3 $\mu$K, negligible compared to the rms amplitude of
$\sim 150\, \mu$K
of the intrisic small scale CMB fluctuations.

The input CMB anisotropy spectrum used for the CMB sky model 
was chosen to fit the results of the joint Maxima-Boomerang data 
analysis (Jaffe
et al. 2001): 
$\Omega=1$, $\Lambda=0.7$, $h=.82$, $\Omega_b h^2 = 0.03$ and
$n_s=0.975$.

The number of MC simulations performed to
estimate the noise, the high-pass filter related transfer function,
and the $C_\ell$ statistics was,
respectively, $\nmcn=450$, $\nmcs=250$ and $\nmcsn=1350$. 
Such high numbers of MC simulations are not
necessary in practice, but were executed 
here to test accurately for  possible biases, and to measure
to a good precision the statistical distribution of the
$C_\ell$ estimates.
Three different apodisations of the analysed  sky region,
described in Table~\ref{table:apodisation}, were
used for  power spectrum estimation. 

%-------------------------------------------------------
% \begin{table*}[t]
% \begin{tabular}{cllllll}
% Apodisation & $W(r \leq \rho(\phi))$       & $W(r>\rho(\phi))$ &$w_1$ &
% $w_2$ & $w_4$ & $w_2^2/w_4$ \\
% \tableline
% Top-hat     & $1$        &$0$  &$1$  & $1$  & $1$ & $1$\\
% Gaussian    & $\exp\left({-\frac{1}{2}\left(\frac{3r}{\rho(\phi)}\right)^2}\right)$
% &$0$  &$0.221$ &$ 0.116$ &$ 0.0559$ & $0.223$ \\
% Cosine      & $\cos\left(\frac{\pi\ r}{2\ \rho(\phi)}\right)$ 	 &$0$  & $ 0.464$  & $ 0.298$
% &$ 0.173$ & $0.514$\\
% \hline
% \end{tabular}
% \caption{Apodisation applied to the ellipse when extracting  the power
% spectrum, where $\rho(\phi) = a/\sqrt{1 + (a^2/b^2 - 1) \sin^2\phi}$ is the
% radius of the ellipse in a direction $\phi$, and $w_i = \int d\r
% W(\r)^i/4\pi\fsky$} % when $a=20^o$ and $b=12^o$}
% \label{table:apodisation}
% \end{table*}
\begin{table}[t]
\begin{tabular}{cll}
Apodisation & $W(r \leq \rho(\phi))$        & $w_2^2/w_4$ \\
\tableline
Top-hat     & $1$        & $1$\\
Cosine      & $\cos\left(\frac{\pi\ r}{2\ \rho(\phi)}\right)$ 	 & $0.514$\\
Gaussian    & $\exp\left({-\frac{1}{2}\left(\frac{3r}{\rho(\phi)}\right)^2}\right)$
 & $0.223$ \\
\tableline
\end{tabular}
\caption{Apodisation applied to the elliptically shaped region of the
sky
used for   extraction of  the CMB anisotropy power
spectrum.
$\rho(\phi) = a/\sqrt{1 + (a^2/b^2 - 1) \sin^2\phi}$ is the
radius of the ellipse in a direction $\phi$, 
and $w_i = \int d\n
W(\n)^i/4\pi\fsky$} % when $a=20^o$ and $b=12^o$}
\label{table:apodisation}
\end{table}
%-------------------------------------------------------

%--------------- figure : C_\ell stat --------------------
\begin{figure*}[ht]
  \psfig{bbllx=90pt,bblly=420pt,bburx=530pt,bbury=750pt,angle=0,file=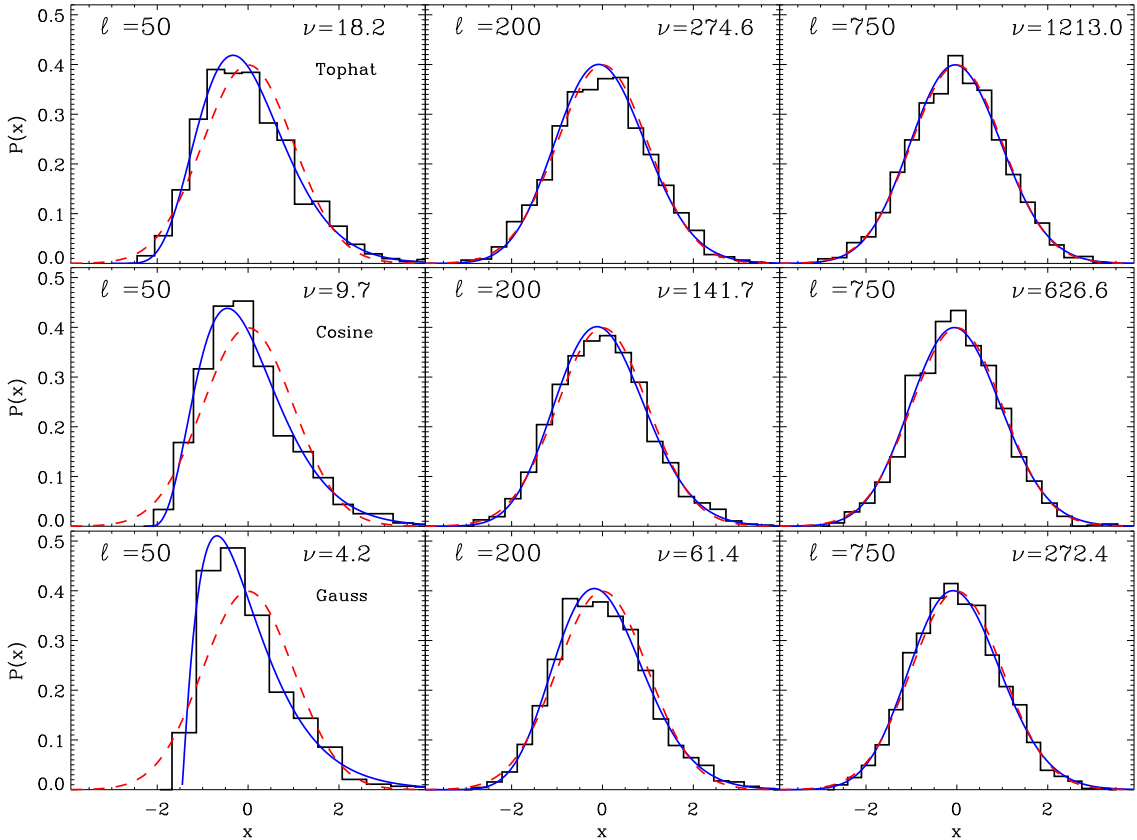}
\caption{Statistical distribution of the MASTER estimates of the binned 
full sky CMB anisotropy power spectrum are shown for the bins centered
at $l=50,\ 200,\ {\rm and}\  750$. The results  of application
of three different sky window apodisations, the Tophat, Cosine, and Gaussian,
are shown in top to  bottom panels, respectively.
The abscissa shows the values of the binned spectrum estimates, 
${\widehat{\mathcal{C}}}_b$, centered and normalised with the
theoretical
values used in the MASTER simulations:
$x = ({\widehat{\mathcal{C}}}_b - {\mathcal{C}}_b^{\rm th})/\Delta
{\widehat{\mathcal{C}}}_b$ (see Eq.~\ref{eq:optim_error}). 
The histogram
shows the distribution derived from the  
MASTER simulations, the (blue) solid line is derived from
the $\chi^2_\nu$ model where $\nu$ is given by Eq.~(\ref{eq:approx_nutrans}), 
and the (red) dashed line a Gaussian of same mean and variance.}
\label{fig:test_mc_histo}
\end{figure*}
%-------------------------------------------------------

\subsection{Statistics of Binned Spectrum Estimates}
\label{section:stat_of_estim}
Figure~\ref{fig:test_mc} shows the Monte Carlo averages and errors  on the 
estimated ${\widehat{\mathcal{C}}}_b$-s, 
computed with the binwidth $\Delta l = 50$ 
for the three used  sky window apodisations. 

Clearly, the average recovered power spectrum
is very close to the input one both at small $\ell$-s,
where the signal is dominant,
and at large $\ell$-s, where the signal to noise ratio is very low.
This demonstrates 
that the estimator (\ref{eq:Cb_estim}) is not biased.
The error bars obtained from the Monte Carlo simulations
(Eq.~\ref{eq:mcsht_error}) can be compared
to the ``naive'' ones (Eq.~\ref{eq:approx_errbar}) corrected for the
transfer function $F_\ell$. This effect decreases 
the effective number of modes  at each
$\ell$ to 
\beq
	\nu_b^f =(2\ell_b+1)\, \Delta \ell\, \fsky \, \frac{w_2^2}{w_4} F_{\ell_b},
	\label{eq:approx_nutrans}
\eeq
which renders the new analytical error estimate
\beq
\Delta  {\widehat{\mathcal{C}}}_b =
	\left({{\mathcal{C}}}_b^{\rm th}+{\widehat{\mathcal{N}}}_b\right)\sqrt{\frac{2}{\nu_b^f}}. \label{eq:optim_error}
\eeq
Figure~\ref{fig:test_mc} shows that the MC error bars are almost
identical to these analytical estimates.
Figure~\ref{fig:test_mc} also 
shows  the spectrum estimate  bin-bin correlation matrix, 
renormalised such that the diagonal
is unity. This matrix is diagonally dominated and except
for the first few bins, the off-diagonal elements 
are at most 10\% as large as the
diagonal elements.

The distribution of the MASTER ${\widehat{\mathcal{C}}}_b$
estimators measured in an $\ell$-bin $b$
defines the likelihood
$P({\widehat{\mathcal{C}}}_b|{\mathcal{C}}^{\rm th})$ of measuring
${\widehat{\mathcal{C}}}_b$ given the input
power spectrum ${\mathcal{C}}^{\rm th}$ (after likelihood marginalisation over
all the other bins). This distribution was estimated from 1352 Monte
Carlo simulations, and is illustrated 
in figure~\ref{fig:test_mc_histo}.
The histograms show the simulated 
distribution together with  the  scaled $\chi^2_{\nu}$ model,
% \beq
% 	P({\widehat{\mathcal{C}}}_b | {\mathcal{C}}_b^{\rm th}) = 
% \chi^2_{\nu}\left(\frac{{\widehat{\mathcal{C}}}_b\ \nu}{ {\mathcal{C}}_b^{\rm
% th}}\right)
% 	\label{eq:chitwo}
% \eeq
where $\nu$ (given by Eq.~\ref{eq:approx_nutrans}) is indicated on each panel,
and a Gaussian distribution with the same mean and variance,
which becomes undistinguishable from the $\chi^2$ curve 
for large $\nu$. At all
$\ell$-s, and specially for small values of $\nu$, the $\chi^2$ model, which 
has no free parameters, gives a much better description of the actual
distribution than the Gaussian.

To estimate the cosmological parameters one seeks a
theoretical model $\Cltheory$ defined by these parameters,
for which the likelihood
$P({\mathcal{C}}^{\rm th}|{\widehat{\mathcal{C}}}_b)$ constructed for
a given data set
is maximised.
Using the Bayes theorem, this can be rewritten as
\begin{equation}
	P({\mathcal{C}}^{\rm th}|{\widehat{\mathcal{C}}}_b) \propto
	P({\widehat{\mathcal{C}}}_b|{\mathcal{C}}^{\rm th})
	P({\mathcal{C}}^{\rm th}),
\end{equation}
where $P({\mathcal{C}}^{\rm th})$ is the prior on the cosmological parameters.
The term $P({\widehat{\mathcal{C}}}_b|{\mathcal{C}}^{\rm th})$ is often
assumed to be a Gaussian function of the data
${\widehat{\mathcal{C}}}_b$, whereas its dependence on the theory
${\mathcal{C}}^{\rm th}$ can be approximated by an offset-lognormal
function  (Bond, Jaffe \& Knox 2000). We see however from
Fig.~\ref{fig:test_mc_histo} that for experiments with small sky
coverage, the likelihood
$P({\widehat{\mathcal{C}}}_b|{\mathcal{C}}^{\rm th})$
has to be described at small $\ell$-s as 
$\chi^2$ function of ${\widehat{\mathcal{C}}}_b$ to avoid biasing the
power spectrum estimation and the cosmological parameters extracted
from it.

\subsection{Monte Carlo Convergence of the Power Spectrum Estimator}

We checked, in the  Monte Carlo simulations  described above,  
that the estimators for the transfer function, the noise
power spectrum (on the sky),
and the accuracy  of the error on ${\widehat{\mathcal{C}}}_b$ 
converge with the number of MC cycles, respectively, as follows 
\beq
	\frac{\delta F_{\ell}}{F_{\ell}} \sim a\; 10^{-2} \sqrt{\frac{100}{\nmcs}}
	\sqrt{\frac{400}{\ell}}
\eeq
with $a$ between 0.6 and 0.9, depending on the sky window used;
\beq
	\frac{\delta {\widehat{\mathcal{N}}}_b}{{\widehat{\mathcal{N}}}_b} \sim0.6\ 10^{-2} \sqrt{\frac{100}{\nmcn}}
	\sqrt{\frac{400}{\ell_b}}\sqrt{\frac{50}{\Delta \ell}},
\eeq
and
\beq
	\frac{\delta(\Delta{\widehat{\mathcal{C}}}_b)}{\Delta{\widehat{\mathcal{C}}}_b } \sim 0.1 \sqrt{\frac{100}{\nmcsn}}.
\label{eq:errMCerr}
\eeq
The contribution of the MC based estimation of $F$ and
${\widehat{\mathcal{N}}}$ to the error on the recovered power spectrum is
\beq
	\frac{\delta
{\widehat{\mathcal{C}}}_b}{{\widehat{\mathcal{C}}}_b} =
	\frac{\delta F_{\ell}}{F_{\ell}} + \frac{\delta
{\widehat{\mathcal{N}}}_b}{{\widehat{\mathcal{N}}}_b} 
	\frac{\widehat{\mathcal{N}}_b}{{\widehat{\mathcal{C}}}_b} 
\eeq
and can be as low as a few percent for $\nmcs \sim \nmcn \sim 100$ in
the signal dominated regime where ${\widehat{\mathcal{N}}}_b \ll
{\widehat{\mathcal{C}}}_b$. In the noise dominated regime, the ratio
of the method induced uncertainty $\delta
{\widehat{\mathcal{C}}}_b$ to the statistical error $\Delta
{\widehat{\mathcal{C}}}_b$ is 
\beq
	\frac{\delta {\widehat{\mathcal{C}}}_b}{\Delta
{\widehat{\mathcal{C}}}_b} \simeq 0.1 \sqrt{\frac{100}{\nmcn}},
\eeq
can be brought below $10\%$ in about 100 Monte Carlo simulations and
analysis of pure noise TOD streams.
Similarly, from Eq.~(\ref{eq:errMCerr}), estimates of the indidual statistical errors of each
binned $\Delta{\widehat{\mathcal{C}}}_b$ better than 10\% can be
obtained with a few hundred MC cycles. We have seen however in section
\ref{section:stat_of_estim} that
it is possible to predict analytically these errors with great accuracy.

\section{Conclusions}
\label{section:conclusion}
We have introduced a MASTER method for rapid estimation of the angular
power spectrum of the CMB anisotropy from the modern CMB data sets.
The method is
based on direct spherical harmonic
transform of the observed area of the sky and   
Monte Carlo simulations of the relevant details of observations 
and data processing.
We demonstrated in an application of the MASTER method to  simulated
observations of \bldb\   that this method renders
unbiased 
estimates of the CMB power spectrum, with error bars very close to optimal.
Monte Carlo calibration of the MASTER method requires generation and analysis
of $\la 1000$  independent simulated realisations of a given CMB experiment. 
We demonstrated that CPU time requirements of the MASTER approach
permit succesful analysis of the largest CMB data sets that exist at
the present time  
using very modest computer facilities, for example an inexpensive PC farm.

Because of the combination of the unsophisticated map making that we used 
and the aggressive
high pass filtering applied to the Boomerang data stream 
in our numerical tests of MASTER, the estimated power spectrum at the lowest
multipoles ($\ell<100$) has relatively large error bars. 
We are currently investigating the use of a more
sophisticated map making algorithm to try to improve this situation.
Ultimately, however, in the large sky coverage experiments, 
such as MAP or Planck, the power spectrum at low $\ell$-s
can be analysed with fully fledged likelihood techniques if a coarsened
pixelisation is used.

Possible improvements of the method include the modelisation of
specific systematic effects, the use of more sophisticated map making
techniques, and the extension to CMB polarisation measurements.

\acknowledgments{EH would like to thank O. Dor\'e for stimulating 
discussions, J. Ruhl for useful comments on the
manuscript, A. Lange for continuous encouragement while this technique 
was developed, and all the Boomerang team for
providing such a stimulating environnement. 
We thank A.J. Banday for help with creating the acronym for our method
and for his careful reading of the manuscript. 
We acknowledge the use of HEALPix, cmbfast and fftw.}
%===========================================================================
%===========================================================================
%===========================================================================
\appendix
\section{Appendix: mode-mode coupling kernel}
In this appendix we compute the mode-mode coupling kernel  resulting
from the cut sky analysis, both in the planar geometry case, where the
arithmetics involved may be more familiar and on the sphere.
\subsection{Analysis on the plane}
A scalar field $\Delta T(\r)$ defined on the plane 
(or defined on the sphere
and projected on a tangent plane) 
can be decomposed in Fourier coefficients as follows
\beq
	a({\k}) = \int d\r \Delta T(\r) e^{-2i\pi\k\r},
\eeq
and
\beq
	\Delta T(\r) = \int d\k a({\k}) e^{2i\pi\k\r}.
\eeq

If $\Delta T$ is the homogeneous, isotropic, 
Gaussian distributed temperature fluctuation, 
each $a(\k)$ is an independent  Gaussian random  variable with
\beq
	\VEV{a({\k})} = 0,
\eeq
and
\beq
	\VEV{a({\k})a^*({\k'})} = \delta (\k-\k')\VEV{C(\k)}= \delta (\k-\k')\VEV{C(k)},
	\label{eq:ortho_ak}
\eeq
where $\delta$ is the Dirac delta function.

The Fourier coefficients derived on a weighted plane are then
\begin{eqnarray}
	\tilde{a}({\k}) &=& \int d\r \Delta T(\r) W(\r) e^{-2i\pi\k\r}, \\
% 	               &=& \int d\k' a({\k'}) \int d\r   e^{2i\pi\k'\r}W(\r) e^{-2i\pi\k\r}, \\
	               &=& \int d\k' a({\k'}) K_{\k'\k}[W]. 
\end{eqnarray}
If we write $W(\r) = \int{d\k} w({\k}) e^{2i\pi\k\r} $, 
the coupling kernel reads
\begin{eqnarray}
	K_{\k_1\k_2} 	&\equiv& \int d\r  e^{2i\pi\k_1\r} W(\r)
			e^{-2i\pi\k_2\r} \\
			& = & \int \k_3 w({\k_3}) \int d\r  e^{2i\pi\r(\k_1-\k_2+\k_3)} \nonumber \\
			& = & \int \k_3 w({\k_3}) \delta(\k_1-\k_2+\k_3).
		\label{eq:kernel_kkk}
\end{eqnarray}
If the polar coordinates of the vector $\k_i$ are 
$(k_i,\theta_i)$, the following useful property of the Dirac delta 
function can be demonstrated:
\begin{eqnarray}
	\int\int {d\theta_1}{d\theta_2}\delta(\k_1+\k_2+\k_3)
	  &=&
	\int{d\theta_2}\delta(k_1-|\k_2+\k_3|)/k_1 \nonumber \\
	&=&  {2\pi}J(k_1, k_2, k_3),
\end{eqnarray}
where the function $J$ is defined as follows
\beq 
	J(k_1, k_2, k_3) = \frac{2}{\pi}\left(2k_1^2k_2^2 + 2k_1^2k_3^2 + 2k_2^2k_3^2 -
	k_1^4 - k_2^4 - k_3^4 \right)^{-1/2}
\eeq
for $|k_2-k_3|<k_1<k_2+k_3$, and $J=0$ otherwise. It follows that
\begin{eqnarray}
	\int  d\k_1 J(k_1, k_2, k_3) &=& \pi \int_{(k_2-k_3)^2}^{(k_2+k_3)^2} d(k_1^2) J(k_1, k_2, k_3) \nonumber \\
	&=& 2 \int_{-1}^{1}
	du(1-u^2)^{-1/2} \nonumber \\
	&=& 2\pi,
	\label{eq:J_integral}
\end{eqnarray}
and  $J(k_1, k_2, 0) = \delta(k_1-k_2)/k_1$.

\begin{eqnarray}
	\VEV{\tC_{k_1}} 
		& \equiv& 
		\frac{1}{2\pi}\int{d\theta_1}\VEV{\tilde{a}({\k_1})\tilde{a}^*({\k_1})}, \\
		& = &
		\frac{1}{2\pi}\int{d\theta_1}\int{d\k_2}\int{d\k_3}
		\VEV{{a}({\k_2}){a}^*({\k_3})}
		K_{\k_1\k_2}[W]K^*_{\k_1\k_3}[W] \nonumber \\
		& = &
		\frac{1}{2\pi}\int{d\theta_1}\int{d\k_2}\VEV{C_{k_2}}\int{d\theta_2}
		\left|K_{\k_1\k_2}[W]\right|^2 \nonumber \\
		& = & 2\pi \int \int k_2 dk_2\, k_3 dk_3\,  
		\VEV{C_{k_2}}\,
		{\mathcal{W}}(k_3)\, J(k_1,k_2,k_3),
		\nonumber
\end{eqnarray}
where ${\mathcal{W}}(k) = \int d\theta  w(\k)w(\k)^* / 2\pi$.

This equation can be rewritten as follows
\beq
\VEV{\tC_{k_1}} = \int k_2 dk_2\, M_{k_1 k_2} \VEV{C_{k_2}},
\eeq
where the coupling kernel is given by 
\beq
 M_{k_1 k_2} = 2 \pi \int k_3 d k_3 {\mathcal{W}}(k_3)\,
J(k_1,k_2,k_3).
\label{eq:kernel_plane}
\eeq

If $C(k_2) = N$ (the white noise spectrum) then
\beq
	\VEV{\tC_{k_1}} = 2\pi  N \int k_3 dk_3 {\mathcal{W}}(k_3),
\eeq
where we used Eq.~(\ref{eq:J_integral}).
%---------------------------------------------------------------
%---------------------------------------------------------------
\subsection{Analysis on the sphere}

A scalar field $\Delta T(\n)$ defined on the sphere and weighted with
an
arbitrary window function $W(\n)$ can be expanded in spherical
harmonics
as follows
\begin{eqnarray}
	\tilde{a}_{\ell m} &=& \int d\n \Delta T(\n) W(\n) Y^*_{\ell m}(\n),\\
	               &=& \sum_{\ell'm'} a_{\ell'm'} \int d\n  Y_{\ell'm'}(\n) W(\n) Y^*_{\ell m}(\n), \\
	               &=& \sum_{\ell'm'} a_{\ell'm'} K_{\ell ml'm'}[W], 
\end{eqnarray}
where the kernel $K$ describes the mode-mode coupling resulting from 
the sky weighting. If $W$ is z-axis azimuthally symmetric, 
$K_{\ell ml'm'} = K_{\ell ml'm} \delta_{mm'}$.
See Wandelt, Hivon \& G\'orski (1999) 
for an analytical calculation of $K_{\ell ml'm}$ in
the case when $W$ is a tophat window.

Note that $\tilde{a}$ is a linear combination of Gaussian variables and is
therefore Gaussian as well, but the $\tilde{a}_{\ell m}$-s 
are not independent.

If we use the series representaion of the window function,
$W(\n) = \sum_{\ell m} w_{\ell m} Y_{\ell m}(\n)$,
the coupling kernel reads
\begin{eqnarray}
	K_{\ell_1m_1\ell_2m_2} 	&\equiv& \int d\n  Y_{\ell_1m_1}(\n) W(\n) Y^*_{\ell_2m_2}(\n) \\
			& = &  \sum_{\ell_3m_3} w_{\ell_3m_3} \int d\n  Y_{\ell_1m_1}(\n)
			Y_{\ell_3m_3}(\n) Y^*_{\ell_2m_2}(\n)   \nonumber \\
			& = & \sum_{\ell_3m_3} w_{\ell_3m_3} (-1)^{m_2}
			\left[ \frac{(2\ell_1+1)(2\ell_2+1)(2\ell_3+1)}{4\pi}
			\right]^{1/2} \nonumber \\
			& \  & \quad \quad \times
			\wjjj{\ell_1}{\ell_2}{\ell_3}{0}{0}{0}
			\wjjj{\ell_1}{\ell_2}{\ell_3}{m_1}{-m_2}{m_3}, 
		\label{eq:kernel_jjj}
\end{eqnarray}
where we introduced the Wigner 3-$j$ symbol (or Clebsch-Gordan coefficient)
$\wjjj{\ell_1}{\ell_2}{\ell_3}{m_1}{m_2}{m_3}$.
Several properties of the 3-$j$ symbol will prove useful.
This scalar object describes  the coupling of 3 angular momentum
vectors (whose squared moduli are $\ell_i(\ell_i+1)$, and
projections on the same axis are $m_i$, for $i=1,2,3$) 
such that the total angular momentum vanishes.
$\wjjj{\ell_1}{\ell_2}{\ell_3}{m_1}{m_2}{m_3}$ 
is non zero only if the triangle relation
\beq
	|\ell_1-\ell_2| \leq \ell_3 \leq \ell_1+\ell_2
	\label{eq:triangle}
\eeq
is satisfied, and 
\beq
	m_1+m_2+m_3 = 0.
	\label{eq:m_wigner}
\eeq
The orthogonality relations of the Wigner symbols read
\beq
	\sum_{\ell_3m_3} (2 \ell_3+1)
	\wjjj{\ell_1}{\ell_2}{\ell_3}{m_1}{m_2}{m_3}
\wjjj{\ell_1}{\ell_2}{\ell_3}{m_1'}{m_2'}{m_3}
	= 
	\delta_{m_1m_1'}\delta_{m_2m_2'},
	\label{eq:ortho_wigner1}
\eeq
\beq
	\sum_{m_1m_2} \wjjj{\ell_1}{\ell_2}{\ell_3}{m_1}{m_2}{m_3}
	\wjjj{\ell_1}{\ell_2}{\ell_3'}{m_1}{m_2}{m_3'} =	
	\delta_{\ell_3\ell_3'}\delta_{m_3m_3'}
\delta(\ell_1,\ell_2,\ell_3) \frac{1}{2\ell_3+1},
	\label{eq:ortho_wigner2}
\eeq
where $\delta(\ell_1,\ell_2,\ell_3) = 1$ when the triangular relation
(\ref{eq:triangle}) is satisfied,  and 
$\delta(\ell_1,\ell_2,\ell_3) = 0$
otherwise. Finally, several recursive or closed form  relations
can be used to compute the $3-j$ symbols. A useful example of the
latter is 
\beq
	\label{eq:express_wigner}
	\wjjj{\ell_1}{\ell_2}{\ell_3}{0}{0}{0} = (-1)^{L/2}
\left[\frac{(L-2\ell_1)!(L-2\ell_2)!(L-2\ell_3)!}{(L+1)!}\right]^{1/2}
	\frac{(L/2)!}{(L/2-\ell_1)!(L/2-\ell_2)!(L/2-\ell_3)!}
\eeq
for even $L\equiv \ell_1+\ell_2+\ell_3$ (or equal to 0 for odd $L$),
with the asymptotic behaviour for $L \gg 1$
\beq
%%%%%	\wjjj{\ell_1}{\ell_2}{\ell_3}{0}{0}{0}^2 \longrightarrow \frac{2}{\pi}\left(L(L-2\ell_1)(L-2\ell_2)(L-2\ell_3)\right)^{-1/2}.
	\wjjj{\ell_1}{\ell_2}{\ell_3}{0}{0}{0}^2 \longrightarrow
\frac{2}{\pi}\left(2\ell_1^2\ell_2^2+2\ell_1^2\ell_3^2+2\ell_2^2\ell_3^2
- \ell_1^4 - \ell_2^4 - \ell_3^4\right)^{-1/2}.
\eeq
See Edmonds (1957) for further details on Wigner symbols.

The ensemble averaged power spectrum of the random scalar field 
$\Delta T(\n)$ on the sphere computed with an
arbitrary
weighting function $W(\n)$ can be represented as follows 
\begin{eqnarray}
	\VEV{\tC_{\ell_1}} 
		& \equiv& 
		\frac{1}{2\ell_1+1}\sum_{m_1=-\ell_1}^{\ell_1}\VEV{\tilde{a}_{\ell_1m_1}\tilde{a}^*_{\ell_1m_1}}, \\
		& = &
		\frac{1}{2\ell_1+1}\sum_{m_1=-\ell_1}^{\ell_1}\sum_{\ell_2m_2}\sum_{\ell_3m_3}
		\VEV{{a}_{\ell_2m_2}{a}^*_{\ell_3m_3}}
		K_{\ell_1m_1\ell_2m_2}[W]K^*_{\ell_1m_1\ell_3m_3}[W] \nonumber \\
		& = &
		\frac{1}{2\ell_1+1}\sum_{m_1=-\ell_1}^{\ell_1}\sum_{\ell_2}\VEV{C_{\ell_2}}\sum_{m_2=-\ell_2}^{\ell_2}
		\left|K_{\ell_1m_1\ell_2m_2}[W]\right|^2.
\end{eqnarray}
Upon substituting the kernel expansion in terms of Wigner symbols
(\ref{eq:kernel_jjj}), and reordering the sums, this expression expands to
\begin{eqnarray}
	\VEV{\tC_{\ell_1}} 
		&=&
		\sum_{\ell_2}\VEV{C_{\ell_2}} \frac{2\ell_2+1}{4\pi}
		\sum_{\ell_3m_3}\sum_{\ell_4m_4} w_{\ell_3m_3}w^*_{\ell_4m_4}
		\left( (2\ell_3+1)(2\ell_4+1) \right)^{1/2} \nonumber \\
		&\ & 
		\times\wjjj{\ell_1}{\ell_2}{\ell_3}{0}{0}{0}\wjjj{\ell_1}{\ell_2}{\ell_4}{0}{0}{0}
		\sum_{m_1m_2} \wjjj{\ell_1}{\ell_2}{\ell_3}{m_1}{-m_2}{m_3}\wjjj{\ell_1}{\ell_2}{\ell_4}{m_1}{-m_2}{m_4},
\end{eqnarray}
which can be remarkably simplified with the aid of  both
the orthogonality relation of the
Wigner symbols
(\ref{eq:ortho_wigner2}), 
and the definition (Eq.~\ref{eq:power_window})of the power spectrum
of the window function, ${\mathcal{W}}_{\ell}$.
The final expression reads
\beq
	\VEV{\tC_{\ell_1}} = \sum_{\ell_2} M_{\ell_1\ell_2} \VEV{C_{\ell_2}},
	\label{eq:coupling_final}
\eeq
with
\beq
	M_{\ell_1\ell_2} = \frac{2\ell_2+1}{4\pi}\sum_{\ell_3}
	(2\ell_3+1) {\mathcal{W}}_{\ell_3} \wjjj{\ell_1}{\ell_2}{\ell_3}{0}{0}{0}^2.
	\label{eq:kernel_final}
\eeq
The Wigner symbols can be numerically computed from
(\ref{eq:express_wigner}) or from any equivalent recurrence relations.
The equation (\ref{eq:coupling_final}) expresses the ensemble averaged 
angular power spectrum measured with
an {\em arbitrary} window on the sky 
for the statistically  homogeneous and isotropic fluctuations
described by  an arbitrary ensemble
averaged power spectrum over the full sky.

If the input power spectrum is constant, $\VEV{C_{\ell}} = N$, 
corresponding to the white noise distributed over the sky, 
Eq.~(\ref{eq:coupling_final}) can be simplified using
the orthogonality relation (\ref{eq:ortho_wigner1}), and the measured 
windowed power
spectrum is also a constant
\beq
	\VEV{\tC_{\ell}}  = N \fsky w_2.
\eeq

%----------------------------------------------------------
%----------------------------------------------------------
\section{Appendix: transfer function for parallel scans}
\label{append:transfer}
In the case of Boomerang, or of any scanning survey, a crude estimate
of the transfer function corresponding to the high-pass filtering of
the TOD, $\fzero$ (see the Eq.~\ref{eq:transfer_iter}),
can be
obtained by assuming that to first order the scans are parallel 
and performed at
uniform angular speed. 
In such a case the filtering of the TOD will alter the sky signal
structures parallel to scan direction, but leave unchanged the structures
orthogonal to the scan direction. 
If the  surveyed area is small enough ($\sim 20$ deg in each direction) 
the  tangent plane approach is sufficient to model the survey.
Hence, the map can be decomposed in plane waves
\beq
	\Delta T(x,y) = \int a(k_x,k_y) e^{i(k_x x + k_y y)}
\eeq
with $k_x = k \cos\theta$ and $k_y = k\sin\theta$.
The $a(\k)$ are zero-mean Gaussian variables with a variance
\beq
	\VEV{a(\k)a(\k')^*} = \VEV{P(k)} \delta(\k-\k'),  \nonumber
\eeq
and the map power spectrum is given by
\beq
	P(k) = \frac{1}{2\pi} \int d\theta \ a(\k)a(\k)^*.  \nonumber
\eeq
If the scan is performed along the x-axis the map obtained from the
filtered TOD is
\beq
	\Delta T_{\rm filt}(x,y) = 
\int a(k_x,k_y) f(k_x) e^{i(k_x x + k_y y)},  \nonumber
\eeq
and its power spectrum is
\beq
	P_{\rm filt}(k) = \frac{1}{2\pi} \int d\theta a(\k)a(\k)^*
f(k_x)^2.
\eeq
The ensemble  averaged filtered power spectrum  is
\begin{equation}
	\VEV{P_{\rm filt}(k)} = \VEV{P(k)} \frac{1}{2\pi} \int d\theta f(k\cos\theta)^2.
\end{equation}
Hence, the effect of the TOD filtering on the power spectrum of the
map 
amounts to a simple transfer function given by
($\ell \sim k$)
\beq
	\fzerol = 
\frac{1}{2\pi}\int d\theta f(k \cos\theta)^2 \label{eq:approx_transfer}.
\eeq

If the scan is performed at an azimutal speed $v_{az}$ at an elevation
$\theta_{el}$, and the high pass filter applied to the data has a Gaussian
form
\beq 
	f(\nu) = 1 - e^{-(\nu/\nu_c)^2/2}
\eeq
then
\beq
	\fzerol = 2/\pi\ \int_0^{\pi/2} d\theta \big(1-e^{-(\ell\cos\theta/\ell_c)^2/2}\big)^2,
\eeq 
where $\ell_c = 2 \pi \nu_c / (v_{az} \cos \theta_{el})$.
The  asymptotic forms of $\fzerol$ are 
\begin{eqnarray}
	\fzerol &\rightarrow&  \frac{3}{32} \left( \frac{\ell}{\ell_c}
	\right)^4, \quad \ell \ll \ell_c \\
	\fzerol &\rightarrow& 1- \frac{\sqrt{8}-1}{\sqrt{\pi}} \frac{\ell_c}{\ell}, \quad \ell \gg \ell_c.
\end{eqnarray}
On the other hand, if the high pass filter is a chosen in the form of
a sharp cut  at the
frequency $\nu_c$
\begin{eqnarray}
	f(\nu) &=& 0, \quad \nu < \nu_c,  \\
	f(\nu) &=& 1, \quad \nu \ge \nu_c,
\end{eqnarray}
then
\begin{eqnarray}
	\fzerol & = &  0, \quad \ell < \ell_c    
\label{eq:approx_tr_bw_low},\\
	\fzerol & = &  1 - 2/\pi\ \sin^{-1}(\ell_c/\ell), 
\quad \ell \ge \ell_c \label{eq:approx_tr_bw_hi}.
\end{eqnarray}

% If $W(\r)$ is azymutally symmetric, $W(\r) =  \sum_{\ell_3} W_{\ell_3} \int d\r
% Y_{\ell_30}(\r)$ and
% \begin{eqnarray}
% 	K_{\ell_1m_1\ell_2m_2} 	&=& \delta_{m_1m_2} \sum_{\ell_3} W_{\ell_3} (-1)^{m_2}
% 			\left[ \frac{(2\ell_1+1)(2\ell_2+1)(2\ell_3+1)}{4\pi}
% 			\right]^{1/2} \\
% 			& \  & \quad \quad \wjjj{\ell_1}{\ell_2}{\ell_3}{0}{0}{0} \wjjj{\ell_1}{\ell_2}{\ell_3}{m_1}{-m_1}{0} 
% \end{eqnarray}

%======================================================
%======================================================
%======================================================
%======================================================

\end{document}